\documentclass[12pt,a4paper]{article}
\pdfoutput=1
\usepackage[a4paper,total={170mm, 247mm}]{geometry}

\usepackage{amsmath,amsfonts,amssymb}
\usepackage{physics}
\usepackage{bbm}
\usepackage{appendix}
\usepackage{caption}
\usepackage{cite}
\usepackage{color}
\usepackage{datetime}
\usepackage{float}
\usepackage{graphicx}
\usepackage[colorlinks,citecolor=blue,linkcolor=black]{hyperref}
\usepackage[numbers,square,comma,sort&compress,merge]{natbib} 
\usepackage{subfig}
\usepackage{epsfig}
\usepackage{ dsfont }
\usepackage{mathrsfs}

\usepackage{makeidx}

\usepackage{epsf}

\usepackage{slashed}

\usepackage{mathtools}

\usepackage{xcolor}

\usepackage{verbatim}

\def\const{\mathrm{const}}

\def\D{\Delta}

\def\eps{\varepsilon}
\def\f{\frac}

\def\l{\left}

\def\<{\langle}
\def\>{\rangle}
\def\Ocal{\mathcal{O}}
\def\Qcal{\mathcal{Q}}
\def\Ecal{\mathcal{E}}

\def\mc{\mathcal}

\def\m{\mu}

\def\p{\partial}

\def\r{\right}

\def\t{\theta}

\def\be{\begin{equation}}
\def\ee{\end{equation}}

\def\bea{\begin{eqnarray}}
\def\eea{\end{eqnarray}}

\def\ba{\begin{array}}
\def\ea{\end{array}}

\def\bc{\begin{center}}
\def\ec{\end{center}}

\def\bl{\begin{flushleft}}
\def\el{\end{flushleft}}

\def\br{\begin{flushright}}
\def\er{\end{flushright}}

\def\bi{\begin{itemize}}
\def\ei{\end{itemize}}

\def\bt{\begin{tabular}}
\def\et{\end{tabular}}

\definecolor{ao(english)}{rgb}{0.0, 0.5, 0.0}

\newcommand{\REF}[1]{(\ref{#1})}

\numberwithin{equation}{section}

\begin{document}

\begin{titlepage}

\setcounter{page}{0}

\begin{center}

\vspace*{1cm}

{\bf \Large Flux Correlators and Semiclassics}

\vspace{0.5cm}

{\bf Eren Firat$^a$, Alexander Monin$^{a,b}$, Riccardo Rattazzi$^a$, Matthew T.\ Walters$^{a,c,d,e}$}

\vspace{0.5cm}

{\it $^a$ Theoretical Particle Physics Laboratory (LPTP), Institute of Physics, \\
\'{E}cole Polytechnique F\'{e}d\'{e}rale de Lausanne (EPFL), CH-1015 Lausanne, Switzerland} \\
\vspace{1mm}
{\it $^b$ Department of Physics and Astronomy, \\
University of South Carolina, Columbia, SC 29208, USA} \\
\vspace{1mm}
{\it $^c$ Department of Theoretical Physics, \\
Universit\'{e} de Gen\`{e}ve, CH-1211 Gen\`{e}ve, Switzerland}\\
\vspace{1mm}
{\it $^d$ Hamilton Mathematics Institute, School of Mathematics, \\
Trinity College, Dublin 2, Ireland}\\
\vspace{1mm}
{\it $^e$ Maxwell Institute for Mathematical Sciences, Department of Mathematics, \\
Heriot-Watt University, Edinburgh EH14, UK}\\

\end{center}

\vspace{1cm}

\begin{abstract}

We consider  correlators for the flux of energy and charge in the background of operators with large global $U(1)$ charge in conformal field theory (CFT). It has recently been shown that the corresponding Euclidean correlators generically  admit a semiclassical description in terms of the  effective field theory (EFT) for a conformal superfluid. We adapt the semiclassical description to Lorentzian observables and compute the leading large charge behavior of the flux correlators in general $U(1)$ symmetric  CFTs. We discuss the regime of validity of the large charge  EFT  for these Lorentzian observables and the subtleties in extending the EFT approach to subleading corrections. We also consider the Wilson-Fisher fixed point in $d=4-\epsilon$ dimensions, which offers a specific weakly coupled  realization of the general setup, where
the subleading corrections can be systematically computed without relying on an EFT.
\end{abstract}

\end{titlepage}

\tableofcontents


\section{Introduction}

The fundamental laws of physics, as codified by Quantum Field Theory (QFT), are best deduced by studying the near-vacuum dynamics. The latter consists of processes involving a small number of quanta, like for instance $2\to 2$ transitions. The Standard Model (SM) was indeed constructed on those simple methodological grounds. It is nonetheless awe-inspiring how its very same laws also underlie the vastly more complex macroscopic phenomena one finds in condensed matter, chemistry, or biology.

Key to the variety of macroscopic phenomena and to their effective dynamical laws is, obviously, states involving many quanta and, as such, far from the vacuum. The Standard Model (SM) serves as a perfect example, vividly showcasing the diversity of such states and their often intricate dynamics. In light of this, any situation where the system's behavior can be calculated, bridging the gap between  {\it {the few quanta and  the  many quanta}} regimes\footnote{Or, in the spirit of Anderson \cite{Anderson:1972pca},  across the frontier between  {\it less and  more}.}, becomes conceptually intriguing.

As it turns out, an instance of the desired situation is offered by multilegged amplitudes in weakly coupled QFT.
Broadly, in a QFT with a weak coupling $\lambda\ll 1$, one finds that the perturbative series for $n$-legged amplitudes is controlled  both by $\lambda$ and by $\lambda n$.\footnote{Throughout our discussion $\lambda$ will be a quartic coupling.} In practice the dependence on $n$ originates from combinatoric factors
in the diagrammatics~\cite{Cornwall:1990hh,Goldberg:1990qk}.  For $\lambda n\ll1 $ the standard Feynman diagram expansion works well, but at $\lambda n \gtrsim  O(1)$ it breaks down signaling the onset of a new regime.\footnote{ The naive application of perturbation theory actually breaks down when $\lambda n^2\gtrsim O(1)$, but the logarithm of the amplitude can still be computed perturbatively as long as $\lambda n \ll  1$.}
The ranges $\lambda n\ll 1$ and $\lambda n \gg 1$  then naturally and respectively  define the  {\it {few quanta}} and the {\it {many quanta}} regimes. While these facts were certainly known long before, significant progress on their study  was only made in the early 90's \cite{Voloshin:1992mz,Brown:1992ay,Voloshin:1992nu,Libanov:1994ug,Son:1995wz,Rubakov:1995hq}. In particular, focusing on transitions of the type
 ${\it few}\to n$ for massive scalars  near threshold, it was shown that the regime $n\gg1$ is controlled by a semiclassical expansion  \cite{Son:1995wz,Rubakov:1995hq} around a non-trivial complex (and singular) solution, whose ``distance'' from the vacuum  is controlled by the critical parameter $\lambda n$. While this realization helped sort out crucial structural aspects, 
 the properties of the saddle solution at $\lambda n\gg1$ proved hard to tackle. Luckily, the computational  difficulties have been  numerically tackled in a recent remarkable paper \cite{Demidov:2023dim}. There it was explicitly shown that the probability for the ${\it few}\to n$ transition is exponentially suppressed as $n\to \infty$ for boost factors $\gamma\equiv E/m$ of the particles  in the range $1$ to $\sim \! 5$. The results of~\cite{Demidov:2023dim} thus represent one example where,  for energies ranging from non-relativistic to moderately relativistic,
  the dynamics across the frontier between  {\it{less and more}}  is both understood, in terms of a semiclassical expansion, and computed, by numerical methods.

The preceding success story prompts the exploration of other scenarios featuring calculable multilegged amplitudes. Indeed,  thanks to recent progress, one such instance is already offered by  the large charge regime in Conformal Field Theory~\cite{Hellerman:2015nra,Monin:2016jmo,Alvarez-Gaume:2016vff,Cuomo:2017vzg,Banerjee:2017fcx,DeLaFuente:2018uee,Badel:2019oxl,Antipin:2020abu,Badel:2019khk,Arias-Tamargo:2019kfr,Cuomo:2020rgt,Cuomo:2021qws,Cuomo:2021ygt,Cuomo:2021cnb,Dondi:2022wli,Badel:2022fya,Cuomo:2022kio,Cuomo:2023mxg}. There the number of legs $n$ essentially corresponds to a global $U(1)$ charge $Q$.\footnote{Throughout the paper we shall indicate by $Q$ the $U(1)$ charge when dealing with a generic CFT. When specifically dealing with the $U(1)$ Wilson-Fisher model \cite{Badel:2019oxl}, where the charged operator of interest is $\phi^n$ and $Q\equiv n$, we will instead use $n$, to stress this is the number of legs of the operator.} The main result is that the Euclidean correlators involving the insertion of large-$Q$ operators are described by a semiclassical expansion around a non-trivial solution, controlled by inverse powers of $Q$. When working on the cylinder the solution corresponds to a superfluid state with charge density $J^0\sim Q$, in such a way that the behavior at large $Q$ is universally and effectively described by the hydrodynamics of phonons. These results have broad validity and they apply to both weakly coupled models, like Wilson-Fisher models or large $N$ ones, and to generic strongly coupled CFTs. 

The weakly coupled cases \cite{DeLaFuente:2018uee,Badel:2019oxl,Antipin:2020abu} offer indeed the opportunity to study the behavior of the system for arbitrary $\lambda n$, thus allowing us to follow  the dynamics  across the   few quanta ($\lambda n \ll 1$) and the many quanta  ($\lambda n\gg 1$) regimes. One finds that 
$\lambda n$ crucially controls the gap of non-phononic modes, such that for $\lambda n\gg 1$ 
the system behaves as a simple  generic superfluid at distances larger than $1/\lambda n$ on the cylinder.
All these calculations can be carried out analytically, thus illuminating the {\it{less to more}} regime change  without the technical complications of the case of multiparticle production.

The simple superfluid behavior that can be explicitly proven in weakly coupled theories is indeed expected at sufficiently large $Q$ for generic CFTs, including strongly coupled ones. This expectation follows  from the hypothesis of semiclassicality and from the simplest choice for  the symmetry of the saddle solution \cite{Monin:2016jmo}. Crucially, for generic CFTs the large $Q$ description must  be interpreted
as an effective one,  valid only at lengths larger  than the inverse gap 
of the non-phononic excitations on the cylinder. This regime translates into specific ranges for the conformal cross-ratios of operator positions.

The large charge  CFT regime has, so far, been explored in Euclidean signature, or, equivalently, at spacelike separation. The present paper aims to extend the study  to inherently Lorentzian observables, which are closer to those that can be directly measured in collider experiments. That would provide a novel (intrinsically ultra-relativistic) situation in which to study the {\it{less to more}} transition, adding to the above mentioned examples.

\begin{figure}[t!]
    \centering
    \includegraphics[width=0.25\textwidth]{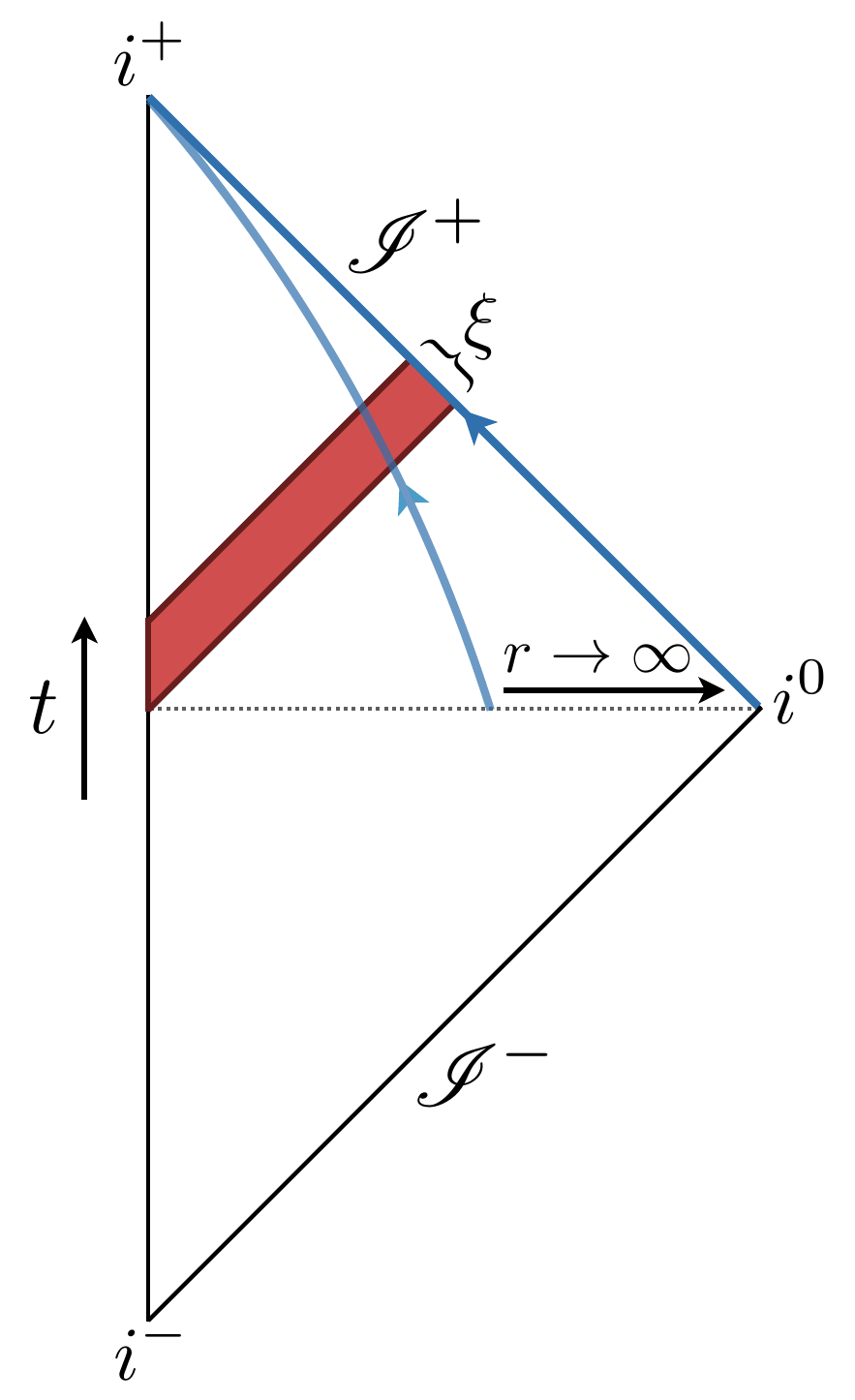}
    \caption{Penrose diagram showing our setup for a spacetime slice of fixed direction $\mathbf{n}$. The operator $\Ocal_Q$ creates a localized excitation at time $t=0$, which propagates radially outwards (red band). The resulting charge and energy flux in a given direction are measured by the line operators~\eqref{eqn: def charge flux x} and \eqref{eqn: def energy flux x}, which in the limit $r \to \infty$ become integrals over future null infinity (blue line). The distribution of this flux in time is set by the typical scale $\xi$ of the excitation.}
    \label{fig: setup}
\end{figure}

One natural set of Lorentzian observables for CFTs was presented in \cite{Hofman:2008ar}: the charge and energy flux resulting from a localized excitation created by a scalar operator. 
In this work, we consider the same setup as~\cite{Hofman:2008ar}, shown in Figure~\ref{fig: setup}, for general CFTs with a global $U(1)$ symmetry, with the local excitation at time $t=0$ created by the lowest-dimension operator $\mathcal{O}_Q$ with charge $Q \to \infty$. We can then place a detector a large distance away in order to measure the total charge flux in a given direction resulting from this excitation,
 \begin{equation}\label{eqn: def charge flux x}
    \mathcal{Q}(\mathbf{n}) \equiv \lim_{r\to\infty} r^{d-2} \int_0^{\infty} dt \, n^i J^i(t,r\mathbf{n}),
\end{equation}
where $J^\mu$ is the conserved current associated with the $U(1)$ symmetry, and the spatial unit vector $\mathbf{n}$ specifies the direction in which we measure the flux. Similarly, we can measure the energy flux
\begin{equation}\label{eqn: def energy flux x}
    \mathcal{E}(\mathbf{n}) \equiv \lim_{r\to \infty} r^{d-2} \int_0^{\infty} dt  \, n^{i}T^{0i}(t,r\mathbf{n}),
\end{equation}
where $T^{\mu\nu}$ is the stress-energy tensor.

Following \cite{Hofman:2008ar} we  consider a state of the form 
\begin{equation}\label{eqn: state}
 \mathcal{O}_Q(E)\>\equiv \int d^dx \, e^{-i E x^0} e^{-\frac{x_E^2}{\xi^2}} \mathcal{O}_Q(x)  |0\rangle,
\end{equation}
where $x_E^2\equiv \sum_\mu (x^\mu)^2$ is a Euclidean norm ensuring the state is normalizable and localized in a region of spacetime where all coordinates are less than $O(\xi)$. In the limit $\xi E\gg 1$ the state approaches a momentum eigenstate, with $P^\mu=(E,\vec 0)$. The idea is that the limit $\xi\to \infty$ is taken after
the limit $r\to \infty$ in eqs.~\eqref{eqn: def charge flux x} and \eqref{eqn: def energy flux x}.

In such a state, we compute correlators with multiple insertions of the flux operators\footnote{These correlators were first investigated by Korchemsky and Sterman~\cite{Korchemsky:1999kt} in the study of QCD jets by relating them to event shape distributions in $e^+e^-$ annihilation, and there has recently been renewed interest in studying their properties in conformal field theories~\cite{Belitsky:2013xxa,Belitsky:2013bja,Belitsky:2013ofa,Hartman:2016lgu,Cordova:2017zej,Kravchuk:2018htv,Belin:2019mnx,Kologlu:2019bco,Kologlu:2019mfz,Huang:2020ycs,Chang:2020qpj,Belin:2020lsr,Korchemsky:2021okt,Korchemsky:2021htm,Chicherin:2023gxt}. The semiclassical approach we use here is complementary to these previous works, which have largely focused on the structure of the light-ray OPE or correlators in weakly coupled or holographic theories. In light of the energies reached by the LHC, it seems that the behavior of flux correlators in CFTs can also have direct relevance to collider physics, as was shown in recent analyses of jet substructure in QCD~\cite{Komiske:2022enw,Lee:2022ige}.}
\begin{equation}\label{eqn: correlators definition}
    \langle \mathcal{Q}(\mathbf{n}_1) \mspace{1mu} \cdots \mspace{1mu} \mathcal{Q}(\mathbf{n}_k)\rangle \equiv \frac{\< \overline{\mathcal{O}}_Q(E) \mspace{1mu} \mathcal{Q}(\mathbf{n}_1) \mspace{1mu} \cdots \mspace{1mu} \mathcal{Q}(\mathbf{n}_k) \mspace{1mu} \mathcal{O}_Q(E)\>}{\< \overline{\mathcal{O}}_Q(E) \mspace{1mu} \mathcal{O}_Q(E)\>},
\end{equation}
with an analogous expression for the energy flux. Conceptually, we can think of the excitation created by $\mathcal{O}_Q(x)$ as roughly $Q$  charged quanta initially localized at $x$, which propagate outwards and are measured at infinity. By measuring the correlation between charge and energy depositions in different directions, we can determine the makeup and dynamics of these charged quanta, just like in a collider experiment.

The overall normalization of these correlation functions is fixed by Ward identities, but in principle they can otherwise have arbitrary dependence on the angular separations $\cos\theta_{ij} = \mathbf{n}_i \cdot \mathbf{n}_j$ between the flux operator insertions. For example, we can write the correlator with two charge flux insertions in the general form
\begin{equation}
    \langle\mathcal{Q}(\mathbf{n}_1) \mspace{1mu} \mathcal{Q}(\mathbf{n}_2) \rangle \equiv \frac{Q^2}{\Omega^2_{d-2}} \big(1+h_Q(d,\theta) \big),
\end{equation}
where $\Omega_{d-2}\equiv\frac{2\pi^{(d-1)/2}}{\Gamma((d-1)/2)}$ is the area of the celestial sphere, and $\theta \equiv \theta_{12}$. Charge conservation imposes the constraint
\begin{equation}\label{inth0}
    \int d\Omega_{d-2} \, h_Q(d,\theta)=0.
\end{equation}

For $Q \gg 1$ in a generic CFT in $d$ spacetime dimensions, the EFT results of~\cite{Hellerman:2015nra,Monin:2016jmo} together with, as we shall argue,  an additional  hypothesis of smoothness of the 4-point function, imply that  $h_Q(d,\theta)$ can be expanded in inverse powers of $Q$ as
\be
\label{eq:Hexp}
h_Q(d,\theta) = h^{(0)}(d,\theta) + \frac{1}{Q^{\alpha}} h^{(1)}(d,\theta) + \ldots
\ee
To be more precise, within the domain of validity of the  EFT of~\cite{Hellerman:2015nra,Monin:2016jmo}, the subleading correction to the 4-point function $\langle \bar\Ocal_Q J^\mu J^\nu\Ocal_Q\rangle$ scales like $Q^{-d/(d-1)}$. However the  flux correlators of eq.~(\ref{eqn: correlators definition}) entail integration over a small portion of coordinate space where the EFT description breaks down. As we discuss in Section~\ref{sec: EFT breaking}, the contribution from this region is still expected to vanish as $Q\to \infty$ under a plausible hypothesis of smoothness. In principle the  resulting leading correction could then scale like  $Q^{-\alpha}$ with $\alpha\not = \frac{d}{d-1}$.

For weakly coupled CFTs, the semiclassical computation of $h_Q$  at large $Q$ (again here we set $Q\equiv n$) should take the same structure as for all other observables~\cite{Badel:2019oxl,Badel:2019khk}. Focusing for definiteness on the  $U(1)$ invariant Wilson-Fisher model in  $d=4-\epsilon$, where the external charged operator is simply $\phi^n$,  we then have
\begin{equation}\label{eq:hWilsonFisher}
\begin{aligned}
h_Q(d,\theta)\to h_n(4\!-\!\epsilon, \lambda_* n, \theta)&=\hat h^{(0)}(\epsilon n,\theta)+\epsilon \mspace{1mu} \hat h^{(1)}(\epsilon n,\theta)+\epsilon^2 \hat h^{(2)}(\epsilon n,\theta)+ \dots\\
&=\hat h^{(0)}(\epsilon n,\theta)+\frac{1}{n}\left [\epsilon n \mspace{1mu} \hat h^{(1)}(\epsilon n,\theta)\right ]+\frac{1}{n^2}\left [(\epsilon n)^2 \hat h^{(2)}(\epsilon n,\theta)\right ]+ \dots
\end{aligned}
\end{equation}
where $\lambda_*= a_1\epsilon+a_2\epsilon^2+\dots$ is the fixed point coupling.
Treating $\lambda_* n\simeq \epsilon n$ as a fixed finite parameter, the expansion in powers of $\epsilon$ can be traded with an expansion in inverse powers of $n$, as shown in the second line. The function associated with each finite order in the $\epsilon$ expansion (equivalently $1/n$ expansion)   describes the transition from the \textit{less} ($\epsilon n\ll 1$) to the \textit{more}  ($\epsilon n\gg 1$) regimes.

In this study, for both the generic  case and the Wilson-Fisher model in $d=4-\epsilon$, we focus on the  leading term in the semiclassical expansion and show that it vanishes
\be\label{eq:h0result}
h^{(0)}(d,\theta) = 0\, ,\qquad \hat h^{(0)}(\epsilon n,\theta)=0\,.
\ee
As shown explicitly in~\cite{Badel:2019oxl,Badel:2019khk}, the dynamics of the Wilson-Fisher model for $n\to \infty$ and  $\epsilon$ fixed (implying $\epsilon n\to \infty$) matches that of the generic EFT. Eqs.~\eqref{eq:Hexp}--\eqref{eq:h0result} are then seen to imply 
\begin{equation}
\lim_{\epsilon n\to \infty} \hat h^{(k)}(\epsilon n,\theta)=0 \quad \forall \, k\,.
\end{equation}
The $k=0$ equation is compatible with eq.~(\ref{eq:h0result}), but the latter result is stronger.

Overall eq.~\eqref{eq:h0result} implies
 homogeneity at $Q\to \infty$. In fact we prove the same result holds for all higher point functions (with $k=$ finite)
\begin{equation}
\label{eq:MainResult}
    \langle \mathcal{Q}(\mathbf{n}_1) \mspace{1mu} \cdots \mspace{1mu} \mathcal{Q}(\mathbf{n}_k)\rangle \to \bigg( \frac{Q}{\Omega_{d-2}} \bigg)^k, \quad \langle \mathcal{E}(\mathbf{n}_1) \mspace{1mu} \cdots \mspace{1mu} \mathcal{E}(\mathbf{n}_k)\rangle \to \bigg( \frac{E}{\Omega_{d-2}} \bigg)^k \qquad (Q \to \infty)\, .
\end{equation}
 This behavior is physically intuitive, since in the limit of an infinite number of outgoing charged quanta we expect the distribution to be isotropic, as was recently argued in~\cite{Chicherin:2023gxt}, which explicitly computed flux correlators in the background of large charge states for the example of $\mathcal{N}=4$ super-Yang-Mills. However, this emergent homogeneity is rather  non-trivial from the perspective of a standard perturbative diagrammatic expansion. Indeed, based on the results of section \ref{sec: free theory} and Appendix \ref{sec: appendix details computation}, one can see that the $O(\lambda_* n)$ term  in the expansion of $\hat h^{(0)}(\epsilon n,\theta)$ vanishes, suggesting that an all-order proof can be systematically constructed.  However  the existence of diagrams scaling like $\lambda_*^p n^q$ with $q>p$, which with careful scrutiny are found to exponentiate and factor out, complicates the construction of such a proof. The same difficulty appears in the diagrammatic computation of the anomalous dimension presented in~\cite{Badel:2019oxl}. In that case, like in this one,\footnote{Or like in the original problem of multiparticle production \cite{Rubakov:1995hq}.} the physics of the problem is properly captured by the semiclassical expansion around the leading saddle point.
That  implies straightforwardly the structure in eq.~(\ref{eq:hWilsonFisher}) as well as $\hat h^{(0)}(\epsilon n,\theta)=0$. Moreover the semiclassical computation extends to the range $\lambda_* n\gg 1$ where the naive diagrammatic approach fails.

In Section~\ref{sec: change of manifold}, we discuss the kinematics and present the general procedure for mapping flux correlators to correlation functions on the Euclidean cylinder. We then consider the charge and energy flux in free field theory in Section~\ref{sec: free theory}, to demonstrate the subtleties involved in their computation. In Section~\ref{sec: semiclassical}, we turn to general CFTs with a $U(1)$ symmetry and use the semiclassical approach to compute the leading order behavior of flux correlators at large charge. We then compute the same observables in the specific example of the Wilson-Fisher fixed point in $d=4-\epsilon$ and discuss the regime of validity of the large charge EFT in Lorentzian signature. Finally, in Section~\ref{sec: discussion} we discuss the possible extension of this framework to subleading corrections and the challenges involved. Various details of our calculations can be found in the appendices.

\section{Kinematics and Coordinate Choices}\label{sec: change of manifold}

In this section, we would like to review some technical aspects, which mostly concern the possible choices of coordinates.

\begin{figure}[t!]
    \centering
    \includegraphics[width=\textwidth]{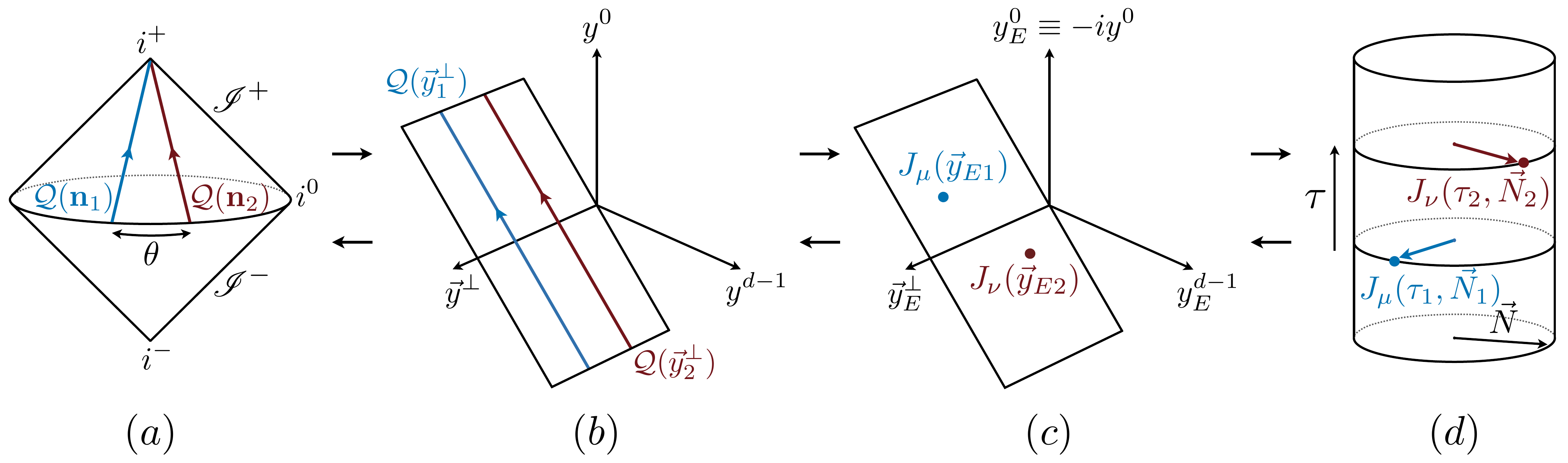}
    \caption{Diagram outlining our procedure for computing flux correlators, for the example of $\<\mathcal{Q} \mathcal{Q}\>$. The flux operator $\mathcal{Q}(\mathbf{n})$ is defined as an integral of the $U(1)$ current over time for a fixed point on the celestial sphere $(a)$. To simplify the computation, the celestial sphere can be mapped to the null plane $y^+=0$ $(b)$. The Lorentzian correlator of the current $J_\mu(y)$ can then be computed by Wick rotating to the Euclidean plane $(c)$ and mapping to the Euclidean cylinder $(d)$, where the semiclassical approach of~\cite{Hellerman:2015nra,Monin:2016jmo} can be used. The result can then be mapped back to the original flux correlator via the same steps in reverse.}
    \label{fig: procedure}
\end{figure}

The definitions of the flux operators in eqs.~\eqref{eqn: def charge flux x}, \eqref{eqn: def energy flux x}  are physically intuitive, but the $r$ and $t$  coordinates are not the most convenient from both the physical and computational point of view. Because of conformal symmetry, and as confirmed by our computation, the state, and with it the conserved charges, spread asymptotically at the speed of light. Quantitatively, this means that at time $t$ in the future, the excitation will be localized around $r\equiv |\vec x|\sim t\pm O(\xi)$, where $\xi $ measures its original distribution (see eq.~\eqref{eqn: state}).
The bulk of the contribution to the integrals in eqs.~\eqref{eqn: def charge flux x} and \eqref{eqn: def energy flux x} then comes from the region $t\sim r \pm O(\xi)$, which goes  to infinity. In this situation, radial light-cone coordinates, consisting of $r^\pm=t\pm r$ and of the radial vector ${\mathbf n}$, provide a more suitable parametrization. Indeed, for any value of $r^+$, the state is localized at finite values of $r^- \sim \xi$. Now, as shown in  Appendix \ref{app:FluxesRelations}, conformal invariance guarantees that not only the total charge, but also the time integrated flux, does not depend on the choice of surface, as long as it approaches future null infinity $\mathscr{I}^+$. The fluxes in eqs.~\eqref{eqn: def charge flux x} and \eqref{eqn: def energy flux x} are then equivalently given by integrals at fixed $r^+$,
\begin{equation}
\label{fluxes-lightcone-coord}
\begin{aligned}
  \mathcal{Q}(\mathbf{n})&= \lim_{r^+\to \infty} \bigg( \frac{r^+}{2} \bigg)^{d-2} \frac{1}{2} \int_{-\infty}^{+\infty} dr^- \, J^{r^+}(r^+,r^-,\mathbf{n}), \\
 \mathcal{E}(\mathbf{n}) &= \lim_{r^+\to \infty} \bigg( \frac{r^+}{2} \bigg)^{d-2} \frac{1}{2} \int_{-\infty}^{+\infty} dr^- \, T^{r^+ r^+}(r^+,r^-,\mathbf{n}),
 \end{aligned}
\end{equation}
as shown in part (a) of Figure~\ref{fig: procedure}, where the integrals are dominated at finite values of $r^-$. The above definitions also show explicitly that the integrated fluxes are light-ray operators~\cite{Belitsky:2013xxa}.

While the radial light-cone coordinates  describe the asymptotic region more conveniently than the original $x^\mu$, they are made slightly inconvenient by their non-trivial metric. Fortunately, following~\cite{Hofman:2008ar}, we can make a conformal transformation $x^\mu \to y^\mu(x)$ to coordinates that share the advantages of radial light-cone coordinates, but not their disadvantages, 
\begin{equation}\label{eqn: y versus x coordinates}
    y^+ = -\frac{1}{x^+}, \qquad y^- = x^- - \frac{|\vec{x}^\perp|^2}{x^+}, \qquad \vec{y}^\perp =\frac{\vec{x}^\perp}{x^+},
\end{equation}
with $x^\pm \equiv x^0 \pm x^{d-1}$, $\vec{x}^\perp \equiv (x^1,\ldots,x^{d-2})$ and similarly for the $y$ coordinates. 
This transformation  maps $\mathscr{I}^+$  to the null plane $y^+=0$~\cite{Hofman:2008ar}.
Parametrizing $x^\mu$ with radial light-cone coordinates one also sees immediately that the surface $r^+=\infty$ (with $r^-$, ${\mathbf n}$ fixed), maps precisely to the plane $y^+=0$. In particular, on this plane  $y^-$ and $\vec{y}^\perp$ are finite functions of $r^-$ and ${\mathbf n}$ 
\begin{equation}
y^-\to \frac{2r^-}{1+n^{d-1}}\,,\qquad \vec{y}^\perp \to \frac{\vec{n}^\perp}{1+n^{d-1}},
\end{equation}
and the null line defined by $r^+=\infty$ and fixed ${\mathbf n}$ maps to the null line defined by $y^+=0$ and fixed $\vec{y}^\perp$. The charge and energy flux operators are then related to the light ray operators defined by integrals on this null line of the currents in the new coordinates,
\be\label{eqn: def flux y}
    \mathcal{Q}(\vec{y}^\perp) \equiv \int_{-\infty}^\infty dy^- \, J_-(y^+=0,y^-,\vec{y}^\perp), \qquad \mathcal{E}(\vec{y}^\perp) \equiv 2 \int_{-\infty}^\infty dy^- \, T_{--}(y^+=0,y^-,\vec{y}^\perp),
\ee
as shown in part $(b)$ of Figure~\ref{fig: procedure}. Accounting for the transformation of the currents and the measure under eq.~\eqref{eqn: y versus x coordinates} the precise relation is\footnote{This relation can be also easily obtained by noting that the total charge $Q$ and total energy $E$ are given by
\be
    Q = \int d\Omega_{d-2} \, \mathcal{Q}(\mathbf{n}) = \int d\vec{y}^\perp \, \mathcal{Q}(\vec{y}^\perp), \qquad E = \int d\Omega_{d-2} \, \mathcal{E}(\mathbf{n}) = \int d\vec{y}^\perp \bigg(\frac{1+|\vec{y}^\perp|^2}{2} \bigg) \, \mathcal{E}(\vec{y}^\perp),
\ee
and identifying the two integration measures $d\Omega_{d-2} = \Big(\frac{1+|\vec{y}^\perp|^2}{2}\Big)^{d-2} d\vec{y}^\perp$ (see Appendix~\ref{app:FluxesRelations} for details).}
\begin{equation}\label{eqn: relation flux x and y}
     \mathcal{Q}(\mathbf{n}) = \bigg(\frac{1+|\vec{y}^\perp|^2}{2}\bigg)^{d-2} \mathcal{Q}(\vec{y}^\perp), \qquad \mathcal{E}(\mathbf{n}) = \bigg(\frac{1+|\vec{y}^\perp|^2}{2}\bigg)^{d-1} \mathcal{E}(\vec{y}^\perp).
\end{equation}
The flux correlators in eq.~\eqref{eqn: correlators definition} are then given by  correlators of light-ray operators on a plane. One key advantage of $y$ coordinates is that any explicit dependence on the diverging coordinate $r^+$ (or equivalently $1/y^+$) is removed. Interested readers may find a more detailed explanation of the relation between  flux operators in different coordinates in Appendix~\ref{app:FluxesRelations}.

We are specifically interested in computing these correlators in the background of operators with large $U(1)$ charge $Q \gg 1$, where we can use the semiclassical approach developed in~\cite{Hellerman:2015nra,Monin:2016jmo}. However, this semiclassical picture is most naturally formulated with radial quantization in Euclidean signature, where such operators create a state of finite charge density, while flux correlators are intrinsically Lorentzian observables. Our strategy will therefore be to Wick rotate the integrands of eq.~\eqref{eqn: def flux y} to the Euclidean plane, $y^0 \to iy_E^0$, as shown in part $(c)$ of Figure~\ref{fig: procedure}. We can then use a Weyl transformation to map the Euclidean plane to the cylinder (part $(d)$ of Figure~\ref{fig: procedure}),
\be\label{eq:WeylTransform}
\vec{y}_E = e^\tau \vec{N},
\ee
where we have set the radius of the cylinder $R=1$. The current and stress tensor correlators on the cylinder are then computed as an expansion around the non-trivial saddle point created by the large charge background.

In short, our calculational procedure is:
\begin{enumerate}
\item Compute the Euclidean correlators
\begin{equation*}
\<Q|\mspace{1mu} J_{\mu_1}(\tau_1,\vec{N}_1) \mspace{1mu} \cdots \mspace{1mu} J_{\mu_k}(\tau_k,\vec{N}_k) \mspace{1mu}|Q\>, \quad \<Q|\mspace{1mu} T_{\mu_1\nu_1}(\tau_1,\vec{N}_1) \mspace{1mu} \cdots \mspace{1mu} T_{\mu_k\nu_k}(\tau_k,\vec{N}_k) \mspace{1mu}|Q\>,
\end{equation*}
on the cylinder using the effective action arising from the finite charge density background, where $|Q\>$ is the ground state in radial quantization with fixed charge $Q$.
\item Map from the cylinder to the plane to obtain the Euclidean correlators
\begin{equation*}
    \<\overline{\mathcal{O}}_Q(\vec{y}_{Ef}) \mspace{1mu} J_{\mu_1}(\vec{y}_{E1}) \mspace{1mu} \cdots \mspace{1mu} J_{\mu_k}(\vec{y}_{Ek}) \mspace{1mu} \mathcal{O}_Q(\vec{y}_{Ei})\>, \quad \<\overline{\mathcal{O}}_Q(\vec{y}_{Ef}) \mspace{1mu} T_{\mu_1\nu_1}(\vec{y}_{E1}) \mspace{1mu} \cdots \mspace{1mu} T_{\mu_k\nu_k}(\vec{y}_{Ek}) \mspace{1mu} \mathcal{O}_Q(\vec{y}_{Ei})\>.
\end{equation*}
\item Wick rotate to Lorentzian signature and integrate over $y_1^-, \, \ldots , \, y_k^-$ to obtain the light-ray correlators
\begin{equation*}
    \<\overline{\mathcal{O}}_Q(y_f) \mspace{1mu} \mathcal{Q}(\vec{y}^\perp_1) \mspace{1mu} \cdots \mspace{1mu} \mathcal{Q}(\vec{y}^\perp_k) \mspace{1mu} \mathcal{O}_Q(y_i)\>, \quad \<\overline{\mathcal{O}}_Q(y_f) \mspace{1mu} \mathcal{E}(\vec{y}^\perp_1) \mspace{1mu} \cdots \mspace{1mu} \mathcal{E}(\vec{y}^\perp_k) \mspace{1mu} \mathcal{O}_Q(y_i)\>.
\end{equation*}
\item Map to the celestial sphere and Fourier transform the positions of the external operators to momentum space to obtain the final correlation functions\footnote{This second operation corresponds to using  the states in eq.~\eqref{eqn: state} and then taking the limit $\xi\to \infty$.}
\begin{equation*}
    \<\mathcal{Q}(\mathbf{n}_1) \mspace{1mu} \cdots \mspace{1mu} \mathcal{Q}(\mathbf{n}_k)\>, \quad \<\mathcal{E}(\mathbf{n}_1) \mspace{1mu} \cdots \mspace{1mu} \mathcal{E}(\mathbf{n}_k)\>.
\end{equation*}
\end{enumerate}


\section{Warmup: Free Scalar}\label{sec: free theory}

Before computing correlators with the semiclassical expansion, it is instructive to first consider the case of a free complex scalar field $\phi$, in order to gain some intuition for the structure of flux correlators and clarify some technical details of our procedure. Here we will focus on the two-point charge flux correlator in the background of the operator $\phi^n$ with $U(1)$ charge $Q=n$,
\begin{equation}\label{eqn:goal free case}
    \< \mathcal{Q}(\mathbf{n}_1) \mspace{1mu} \mathcal{Q}(\mathbf{n}_2) \rangle \equiv \frac{\< \bar{\phi}^n(E) \mspace{1mu} \mathcal{Q}(\mathbf{n}_1) \mspace{1mu} \mathcal{Q}(\mathbf{n}_2) \mspace{1mu} \phi^n(E)\>}{ \<\bar{\phi}^n(E) \mspace{1mu} \phi^n(E)\>},
\end{equation}
as well as the energy flux two-point function $\<\mathcal{E}(\mathbf{n}_1)\mathcal{E}(\mathbf{n}_2)\>$.

As discussed in~\cite{Hofman:2008ar}, the one-point flux correlators in the background of any scalar external source are completely fixed by rotational invariance and the requirement that the integral over all possible directions $\mathbf{n}$ gives the total charge or energy, respectively, leading to the expressions:
\begin{equation}
    \langle \mathcal{Q}(\mathbf{n}) \rangle= \frac{n}{\Omega_{d-2}},\quad \langle \mathcal{E}(\mathbf{n}) \rangle=\frac{E}{\Omega_{d-2}}.
\end{equation}

On the other hand, the two-point correlator can have non-trivial dependence on the angle $\mathbf{n}_1 \cdot \mathbf{n}_2 = \cos\theta$ between the two operators on the celestial sphere. In the case of free field theory, the correlation function can be computed diagrammatically in terms of  the Wightman functions\footnote{The overall factor of $(-1)^{\frac{2-d}{2}}$ arises from Wick rotation from Euclidean to Lorentzian signature and ensures that the one-particle state has positive norm.}, which read 
\begin{equation}\label{eqn:free propagator}
    \<\bar{\phi}(y_1) \mspace{1mu} \phi(y_2)\> = \frac{(-1)^{\frac{2-d}{2}}}{(d-2)\Omega_{d-1}\big((y_{12}^+-i\epsilon_{12}) (y_{12}^- -i \epsilon_{12}) - |\vec{y}_{12}^\perp|^2 \big)^{\frac{d-2}{2}}},
\end{equation}
where $y_{12}^{\mu} \equiv y_1^{\mu} -y_2^{\mu}$ and the infinitesimal $\epsilon_{12} > 0$ ensures proper Wightman ordering, i.e. $\phi(y_2)$ acts before $\bar\phi(y_1)$ (see, e.g.,~\cite{Hartman:2015lfa}).

\begin{figure}[t!]
    \centering
    \includegraphics[width=0.8\textwidth]{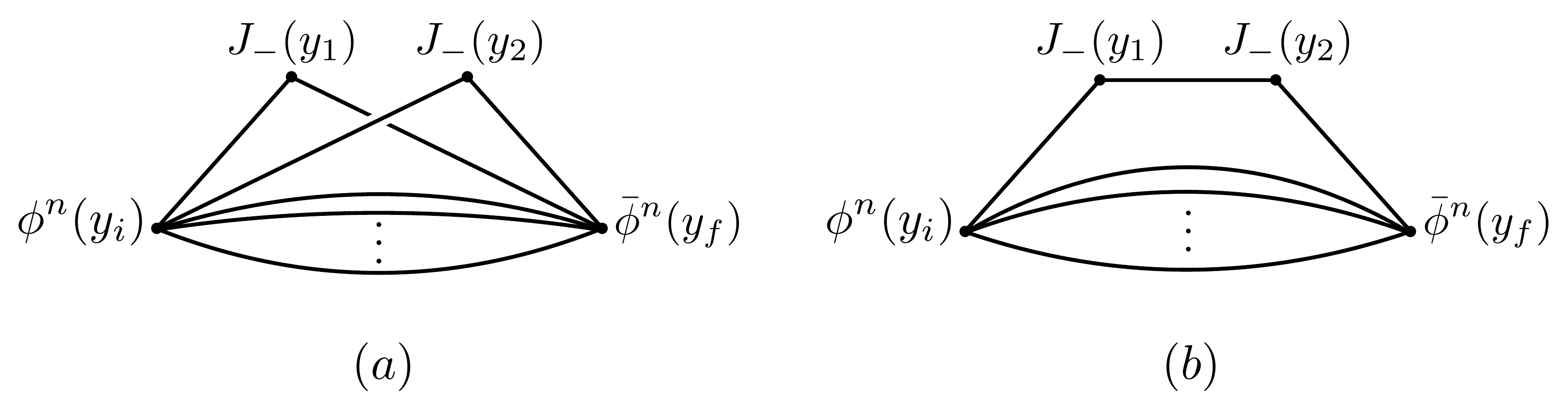}
    \caption{The two types of Wick contractions which contribute to the correlator $\<\bar\phi^n \mathcal{Q} \mathcal{Q} \phi^n\>$ in free theory. The potential contributions from all other Wick contractions vanish upon integrating $y_1^-$ and $y_2^-$. The vertical dots indicate the propagators of the $n-2$ ``spectator'' particles for diagram $(a)$ and the $n-1$ spectators for diagram $(b)$. }
    \label{fig:diagram free case n fields}
\end{figure}

For a free complex scalar field, the $U(1)$ current is given by
\begin{equation}
    J_{\mu}=i \big( \phi (\partial_\mu \bar\phi) - (\partial_\mu\phi) \bar\phi \big),
\end{equation}
which is normalized such that the field $\phi$ has unit charge. To compute the charge flux correlator~\eqref{eqn:goal free case}, we first need the four-point function $\<\bar\phi^n J_- J_- \phi^n\>$. The resulting expression, normalized by the two-point function of $\phi^n$, is
\begin{equation}\label{eq:freeJJ}
    \begin{aligned}
        &\frac{\<\bar\phi^n(y_f) \mspace{1mu} J_-(y_1) \mspace{1mu}  J_-(y_2) \mspace{1mu} \phi^n(y_i)\>}{\<\bar\phi^n(y_f) \mspace{1mu} \phi^n(y_i)\>} \\
        & \quad = (-1)^{d-1} \frac{n(n-1)}{(2\Omega_{d-1})^2} \frac{y_{fi}^{2(d-2)}}{y_{f1}^{d-2} y_{1i}^{d-2} y_{f2}^{d-2} y_{2i}^{d-2}} \bigg( \frac{y_{f1}^+}{y_{f1}^2} + \frac{ y_{1i}^+} {y_{1i}^2}\bigg)  \bigg( \frac{y_{f2}^+} {y_{f2}^2} + \frac{ y_{2i}^+} {y_{2i}^2} \bigg) \\
        & \quad \quad + (-1)^{d-1} \frac{n}{(2\Omega_{d-1})^2} \frac{y_{fi}^{d-2}}{y_{f1}^{d-2} y_{12}^{d-2} y_{2i}^{d-2}} \Bigg( \bigg( \frac{y_{f1}^+} {y_{f1}^2} + \frac{ y_{12}^+} {y_{12}^2} \bigg)  \bigg( \frac{y_{12}^+} {y_{12}^2} + \frac{ y_{2i}^+} {y_{2i}^2} \bigg) + \bigg(\frac{2}{d-2}\bigg) \frac{ (y_{12}^+)^2}{y_{12}^4} \Bigg) \\
        & \quad \quad + \, \ldots,
    \end{aligned}
\end{equation}
where the ``$...$'' stand for contributions to the four-point function that vanish upon integration over $y_1^-$ or $y_2^-$ and therefore do not contribute to the resulting flux correlator, leaving only the two Wick contractions shown in Figure~\ref{fig:diagram free case n fields}. To simplify this expression, we have introduced the shorthand
\be
    y_{jk}^2 \equiv (y_{jk}^+ -i\epsilon_{jk})(y_{jk}^- -i\epsilon_{jk}) - |\vec{y}_{jk}^\perp|^2,
\ee
with $\epsilon_f>\epsilon_1>\epsilon_2>\epsilon_i$.

We then need to integrate over $y_1^-$ and $y_2^-$ to obtain a correlator of charge light-ray operators. Finally, we can use the conformal transformation in eq.~\eqref{eqn: y versus x coordinates} to map to $x$-coordinates and obtain the charge flux correlator,
\begin{equation}\label{eqn: two point function free x space}
    \begin{aligned}
        &\frac{\<\bar\phi^n(x_f) \mspace{1mu} \mathcal{Q}(\mathbf{n}_1) \mspace{1mu} \mathcal{Q}(\mathbf{n}_2) \mspace{1mu} \phi^n(x_i)\>}{\<\bar\phi^n(x_f) \mspace{1mu} \phi^n(x_i)\>} \\
        & \qquad \qquad = \frac{n (n-1)}{\Omega_{d-2}^2} \frac{x_{fi}^{2(d-2)}}{(n_1 \cdot x_{fi})^{d-2} \mspace{1mu} (n_2 \cdot x_{fi})^{d-2}} + \frac{n}{\Omega_{d-2}} \frac{x_{fi}^{d-2}}{(n_1 \cdot x_{fi})^{d-2}} \, \delta(\mathbf{n}_1-\mathbf{n}_2),
    \end{aligned}
\end{equation}
where we have introduced the null vectors $n_j^\mu \equiv (1,\mathbf{n}_j)$, and the delta function is defined such that $\int d\Omega_{d-2} \, \delta(\mathbf{n}-\mathbf{n}') = 1$, for any unit vector $\mathbf{n}'$. The details of this calculation are shown in Appendix~\ref{sec: appendix details computation}.

\begin{figure}[t!]
    \centering
    \includegraphics[scale=0.6]{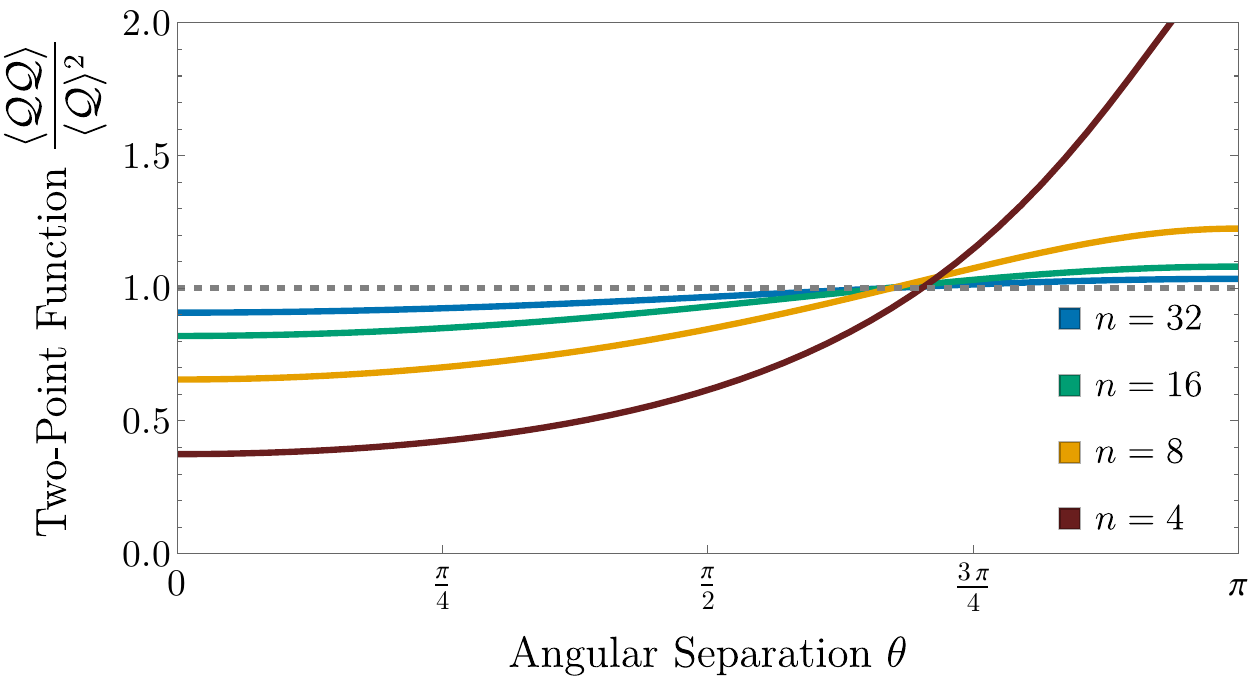}
    \caption{Charge flux two-point correlator $\<\mathcal{Q} \mathcal{Q}\>$ as a function of the angle $\theta$ between the two flux operators in the background of $\phi^4$ (red), $\phi^8$ (gold), $\phi^{16}$ (green), and $\phi^{32}$ (blue), normalized by the homogeneous solution $\<\mathcal{Q}\>^2$. The delta function at $\theta=0$ is not shown.}
    \label{fig:angular dep charge}
\end{figure}

This expression only depends on the separation $x_{fi}$ between the two external sources. Indeed $\mathcal{Q}(\mathbf{n})$ is a translation invariant operator.\footnote{Indeed, for an infinitesimal shift $x^\mu \to x^\mu +a^\mu$, the celestial sphere is invariant: $r^+ $ stays at infinity, while $\mathbf{n}$ is invariant.  Only the time $r^-$ shifts, which is integrated out to compute the flux. Equivalently, in y coordinates,  only $y^-$ transforms under the shift at $y^+=0$, such that the flux is invariant.} Translation invariance of the full correlator implies that it only depends on $x_{fi}$. The two-point function \eqref{eqn:goal free case}'s source has definite energy $E$, and is thus obtained by performing a Fourier transform of $x_{fi}$. After normalizing by the two-point function of $\phi^n$ (see Appendix~\ref{sec: appendix details computation} for details), we obtain:
\begin{equation}\label{eqn: 2 charge correlator result}
    \begin{aligned}
    &\langle \mathcal{Q}(\mathbf{n}_1) \mspace{1mu} \mathcal{Q}(\mathbf{n}_2) \rangle \\
    & \quad = \frac{n(n-1)}{\Omega_{d-2}^2} \frac{\Gamma\big((n-1)(\tfrac{d-2}{2})\big) \Gamma\big(n(\tfrac{d-2}{2})\big)}{\Gamma\big((n-2)(\tfrac{d-2}{2})\big)\Gamma\big((n+1)(\tfrac{d-2}{2})\big)} \, {}_2F_1\Big(d-2,d-2;(n+1)(\tfrac{d-2}{2}); \sin^2\!\tfrac{\theta}{2}\Big) \\
    & \quad \quad + \, \frac{n}{\Omega_{d-2}} \delta(\mathbf{n}_1-\mathbf{n}_2).
    \end{aligned}
\end{equation}
For finite $n$, this correlator has non-trivial dependence on the angle $\theta$ between the two directions on the celestial sphere, as shown in Figure~\ref{fig:angular dep charge}. This dependence is easily understood from momentum conservation: for a source at rest, measuring charge flux in one direction increases the likelihood of measuring flux in the opposite direction. This effect is particularly visible for $n=2$, where eq.~\eqref{eqn: 2 charge correlator result} reduces to a sum of delta functions at $\theta=0$ and $\theta=\pi$. At large $n$, however, this expression converges to the homogeneous distribution
\begin{equation}
    \langle \mathcal{Q}(\mathbf{n}_1) \mspace{1mu} \mathcal{Q}(\mathbf{n}_2) \rangle \approx \bigg(\frac{n}{\Omega_{d-2}}\bigg)^2 \l( 1 + \frac{\Omega_{d-2}}{n} \l [ \delta(\mathbf{n}_1-\mathbf{n}_2)  - \frac{1+ (d-2) \cos \theta }{\Omega_{d-2}} \r ] +O(n^{-2}) \r).
\end{equation}

The fact that the very non-trivial expression \eqref{eqn: 2 charge correlator result} simplifies immensely at leading order in $n$ is understood by noticing that the result can also be found as a saddle point expansion at large $n$ in the path integral. We therefore already see in free theory a first hint that the correct framework to study the large charge regime is a semiclassical expansion.

We can repeat this same calculation for the energy flux operator, with the same diagrammatic contributions as Figure~\ref{fig:diagram free case n fields}. The final result is:
\begin{equation}
    \begin{aligned}
        &\langle \mathcal{E}(\mathbf{n}_1) \mspace{1mu} \mathcal{E}(\mathbf{n}_2) \rangle \\
        &\quad = \frac{E^2}{\Omega_{d-2}^2} \frac{\Gamma\big((n-1)(\tfrac{d-2}{2}) + 1\big) \Gamma\big(n(\tfrac{d-2}{2}) + 1\big)}{\Gamma\big((n-2)(\tfrac{d-2}{2})\big)\Gamma\big((n+1)(\tfrac{d-2}{2})+2\big)} \, {}_2F_1\Big(d-1,d-1;(n+1)(\tfrac{d-2}{2})+2; \sin^2\!\tfrac{\theta}{2}\Big) \\
        & \quad \quad + \, \frac{E^2}{\Omega_{d-2}} \frac{d-1}{n(d-2)+2} \, \delta(\mathbf{n}_1-\mathbf{n}_2).
    \end{aligned}
\end{equation}
The behavior of this expression is analogous to that of the charge flux correlator, converging to the homogeneous distribution $\<\mathcal{E}\>^2$ in the large $n$ limit
\be
\langle \mathcal{E}(\mathbf{n}_1) \mspace{1mu} \mathcal{E}(\mathbf{n}_2) \rangle =  \bigg(\frac{E}{\Omega_{d-2}}\bigg)^2 
\l ( 1 + \f{d-1}{d-2} \f{\Omega_{d-2}}{n} \l [\delta(\mathbf{n}_1-\mathbf{n}_2) - \f{1+(d-1) \cos \t}{\Omega_{d-2}} \r ] + O (n^{-2})\r ). 
\ee


\section{Flux Correlators at Large Charge}\label{sec: semiclassical}

In this section, we compute the charge (and energy) flux correlators~\eqref{eqn: correlators definition} in a semiclassical expansion at large charge. For this, we consider CFTs in $d > 2$ spacetime dimensions with a $U(1)$ global symmetry and we follow the procedure presented in Section~\ref{sec: change of manifold} to compute the correlators in the background of the lowest-dimension operator $\mathcal{O}_Q$ with $U(1)$ charge $Q$. We present the two possible contexts in which we can do the computation. The first possibility is to consider general CFTs with U(1) symmetry and to compute the correlators in the EFT of the Goldstone associated to the U(1) symmetry breaking~\cite{Hellerman:2015nra,Monin:2016jmo}. After briefly reviewing this approach, we apply it to the calculation of the two-point functions $\<\Qcal \Qcal\>$ and $\<\Ecal \Ecal\>$, before generalizing to the case with an arbitrary number of flux operator insertions. While this method has the advantage of being valid for a large class of CFTs, it only investigates the many quanta regime (which in the context of the EFT can be rephrased as the large chemical potential limit $\mu \gg 1$). 
We therefore proceed by showing the computation for a UV-complete theory, the Wilson-Fisher fixed point in $4-\epsilon$ dimensions, where we lose the universality of the result but the semiclassical expansion becomes valid for all values of $\lambda_* n$ (or $\mu$). Finally, we motivate the validity of the EFT to compute the charge and energy flux correlators in Minkowski space.

\subsection{Review of Semiclassics at Large Charge}
\label{semiclassics}

Consider a general Euclidean CFT correlation function of the form
\begin{equation}
\label{planeNpoint}
    \< \overline{\mathcal{O}}_Q(\vec{y}_{Ef}) \mspace{1mu} \Ocal_1(\vec{y}_{E1}) \mspace{1mu} \cdots \mspace{1mu} \Ocal_k(\vec{y}_{Ek}) \mspace{1mu} \mathcal{O}_Q(\vec{y}_{Ei})\>,
\end{equation}
where the intermediate operators $\Ocal_j$ have fixed $U(1)$ charges $q_j \ll Q$ and scaling dimensions $\Delta_j \ll \Delta_Q$ (such as the conserved current $J_\mu$ or stress tensor $T_{\mu\nu}$). Applying conformal transformations, we can always set $y_i\to 0$ and $y_f\to \infty$.
Then, by  Weyl mapping  to  the Euclidean cylinder, we relate eq.~\eqref{planeNpoint}, up to a known rescaling factor, to the cylinder correlator
\begin{equation}
\label{cylNpoint}
    \< Q| \mspace{1mu} \Ocal_1(\tau_1,\vec{N}_1) \mspace{1mu} \cdots \mspace{1mu} \Ocal_k(\tau_k,\vec{N}_k) \mspace{1mu} | Q\>,
\end{equation}
where 
\begin{equation*}
|Q\> \equiv \lim_{\tau\to -\infty} e^{-\Delta_Q\tau} \mathcal{O}_Q(\tau, \vec{N})|0\rangle\,, \qquad\langle Q|\equiv\lim_{\tau\to \infty} e^{\Delta_Q\tau}\langle 0|\mathcal{O}_Q(\tau, \vec{N}),
\end{equation*}
is the ground state  in radial quantization of the conformal multiplet generated by $\mathcal{O}_Q$.
The insight of~\cite{Hellerman:2015nra,Monin:2016jmo} is that, for large  $Q$: (\textit{i}) eq.~\eqref{cylNpoint} can be computed by a systematic expansion  around a saddle point in the path integral; (\textit{ii}) the dynamics around the saddle configuration can be largely determined by symmetry considerations. The simplest option as concerns the second point is that the saddle configuration realizes the spontaneous symmetry breaking
\be \label{eq: breaking pattern}
SO(d+1,1) \times U(1) \to SO(d) \times D',
\ee
where $D'$ is the unbroken combination
\be
D' \equiv D + \mu Q,
\ee
 of dilatations $D$ and of the $U(1)$ charge $Q$, which implements the time translation invariance of the saddle solution. Such a symmetry breaking pattern characterizes the configuration as a conformal superfluid. The parameter $\mu$ has then the standard interpretation of a 
chemical potential (see for instance \cite{Nicolis:2015sra}). Eq.~\eqref{eq: breaking pattern} dictates the presence of (just) one soft Goldstone mode whose properties are largely fixed by symmetry. This is the analogue (with the addition of conformal symmetry) of the hydrodynamic sound mode of ordinary superfluids. The simplest and most natural option  is that this is the only soft mode, while all other modes, which have no symmetry reason to be light, are gapped at the only available mass scale in the system, $\mu$. In several several specific calculable models, this natural set of assumptions has been explicitly verified \cite{DeLaFuente:2018uee,Badel:2019oxl,Badel:2019khk,Antipin:2020abu}.

Under the above hypotheses, the computation of the correlators reduces to a path integral controlled by the effective action for the Goldstone mode. The inserted operators are matched  to local functions of the Goldstone field and its derivatives, precisely like in QCD when mapping quark-gluon operators to mesonic ones.
At leading order in the derivative expansion, the effective Lagrangian is given by 
\be
\label{eq:Leff}
\mathcal{L}_{\textrm{eff}}[\chi] \approx - c_1 (-\partial^\mu\chi \partial_\mu\chi)^{\frac{d}{2}} + \ldots,
\ee
with  $c_1$ is a theory-dependent Wilson coefficient. Setting the  cylinder radius  to $R=1$, the boundary condition corresponding to the state $|Q\rangle$
is implemented by adding the term $i \dot\chi Q/\Omega_{d-1}$ to the Lagrangian~\cite{Monin:2016jmo} in the path integral. The resulting equations of motion
\begin{equation*}
J^0 \equiv \frac{\partial \mathcal{L}_{\textrm{eff}}}{\partial \dot \chi}=i \frac{Q}{\Omega_{d-1}},
\end{equation*}
have the spatially homogeneous solution 
\be\label{eqn: solution on cylinder}
\chi(\tau,\vec{N}) = -i\mu\tau + \textrm{constant},
\ee
with the chemical potential $\mu$ satisfying
\begin{equation}
c_1 \mu^{d-1}\approx \frac{Q}{\Omega_{d-1}}\,\quad \Rightarrow \quad \mu \sim Q^{\frac{1}{d-1}}.
\end{equation}
Replacing the solution back into the action  finally gives the relation between  scaling dimension and charge 
\be
\Delta_Q^{d-1}= \frac{(d-1)^{d-1}}{\Omega_{d-1} d^d} \frac{Q^d}{c_1} \Big( 1 + O(1/Q^{\frac{2}{d-1}}) \Big)\,.
\ee

We can now use the classical solution to compute the leading behavior of correlation functions at large $Q$. To do so, we  match the conserved currents in the correlation functions to the corresponding objects built out of eq.~\eqref{eq:Leff}.
 For example, the $U(1)$ current is given by
\be\label{eq:Jeft}
J_\mu = c_1 d(-\partial^\alpha\chi \partial_\alpha\chi)^{\frac{d-2}{2}} \partial_\mu\chi,
\ee
while the stress tensor is
\be \label{eq:Teft}
T_{\mu\nu} = c_1 d(-\partial^\alpha\chi \partial_\alpha\chi)^{\frac{d-2}{2}} \p_\mu\chi\p_\nu\chi + c_1 (-\partial^\alpha\chi \partial_\alpha\chi)^{\frac{d}{2}} g_{\mu\nu}.
\ee
Using eq.~\eqref{eqn: solution on cylinder} in the above expressions, at leading order in the large $Q$  expansion we then have
\be \label{eq: J and T expectation value}
\begin{aligned}
\<J_\mu\> &\equiv \frac{\<Q| \mspace{1mu} J_\mu(\tau,\vec{N}) \mspace{1mu}|Q\>}{\<Q|Q\>} = -ic_1 d \mu^{d-1} \delta^\tau_{\,\mu} = -\frac{iQ}{\Omega_{d-1}} \delta^\tau_{\,\mu}, \\
\<T_{\mu\nu}\> &\equiv \frac{\<Q| \mspace{1mu} T_{\mu\nu}(\tau,\vec{N}) \mspace{1mu} |Q\>}{\<Q|Q\>} = -c_1 d \mu^d \Big( \delta^\tau_{\,\mu} \delta^\tau_{\,\nu} - \frac{1}{d} g_{\mu\nu} \Big) = -\frac{d}{d-1} \frac{\Delta_Q}{\Omega_{d-1}} \Big( \delta^\tau_{\,\mu} \delta^\tau_{\,\nu} - \frac{1}{d} g_{\mu\nu} \Big).
\end{aligned}
\ee


\subsection{Application to Flux Two-Point Function}\label{sec: flux correlator EFT}

Our goal is to now adapt the semiclassical approach to the calculation of the flux operator correlation functions~\eqref{eqn:goal free case}. Following the procedure laid out in sec.~\ref{sec: change of manifold}, we first need to compute the associated correlator on the Euclidean cylinder using the semiclassical solution~\eqref{eqn: solution on cylinder}. For the current two-point function, we obtain the simple expression,
\be\label{eq:semiclassicJJcyl}
\frac{\<Q| \mspace{1mu} J_\mu(\tau_1,\vec{N}_1) \mspace{1mu} J_\nu(\tau_2,\vec{N}_2) \mspace{1mu} |Q\>}{\<Q|Q\>} = \<J_\mu\> \<J_\nu\> = - \bigg( \frac{Q}{\Omega_{d-1}} \bigg)^2 \delta^\tau_{\,\mu} \delta^{\tau}_{\,\nu}.
\ee
Next, we need to map this correlation function from the cylinder to the Euclidean plane, with all four operators at arbitrary positions. The most general way to do so is to decompose this four-point function into the set of independent tensor structures allowed by conformal symmetry, whose behavior under conformal transformations is known.

Fortunately, because this leading contribution factorizes into a product of CFT two- and three-point functions, which are completely fixed by conformal invariance, we can directly map the individual terms to the Euclidean plane, then Wick rotate to obtain the resulting Lorentzian correlator in $y$-coordinates. Focusing on the only relevant component $J_-$, one then finds 
\be\label{eq:semiclassicJJ}
\begin{aligned}
&\frac{\<\overline{\Ocal}_Q(y_f) \mspace{1mu} J_-(y_1) \mspace{1mu} J_-(y_2) \mspace{1mu} \Ocal_Q(y_i)\>}{\<\overline{\Ocal}_Q(y_f) \mspace{1mu} \Ocal_Q(y_i)\>} = \frac{\<\overline{\Ocal}_Q(y_f) \mspace{1mu} J_-(y_1) \mspace{1mu} \Ocal_Q(y_i)\> \mspace{1mu} \<\overline{\Ocal}_Q(y_f) \mspace{1mu} J_-(y_2) \mspace{1mu} \Ocal_Q(y_i)\>}{\<\overline{\Ocal}_Q(y_f) \mspace{1mu} \Ocal_Q(y_i)\>^2} \\
&\qquad \qquad = (-1)^{d-1} \bigg(\frac{Q}{2\Omega_{d-1}}\bigg)^2 \frac{y_{fi}^{2(d-2)}}{y_{f1}^{d-2} y_{1i}^{d-2} y_{f2}^{d-2} y_{2i}^{d-2}} \bigg( \frac{y_{f1}^+}{y_{f1}^2} + \frac{y_{1i}^+}{y_{1i}^2} \bigg) \bigg( \frac{y_{f2}^+}{y_{f2}^2} + \frac{y_{2i}^+}{y_{2i}^2} \bigg).
\end{aligned}
\ee
The more general analysis in terms of four-point function tensor structures, which is needed for computing subleading corrections in $1/Q$, is shown in appendix~\ref{sec: appendix details computation}.

The position-dependence of the four-point function~\eqref{eq:semiclassicJJ} is identical to the first term in the free theory result from eq.~\eqref{eq:freeJJ}. We can therefore repeat the same analysis to obtain the resulting charge flux two-point function in $x$-coordinates,
\begin{equation}
    \frac{\<\overline{\Ocal}_Q(x_f) \mspace{1mu} \mathcal{Q}(\mathbf{n}_1) \mspace{1mu} \mathcal{Q}(\mathbf{n}_2) \mspace{1mu} \Ocal_Q(x_i)\>}{\<\overline{\Ocal}_Q(x_f) \mspace{1mu} \Ocal_Q(x_i)\>} = \bigg(\frac{Q}{\Omega_{d-2}}\bigg)^2 \frac{x_{fi}^{2(d-2)}}{(n_1 \cdot x_{fi})^{d-2} \, (n_2 \cdot x_{fi})^{d-2}}.
\end{equation}
The charge flux two-point function for a source with definite momentum is then found the same way as for the free scalar field, though in this case, we can only keep the leading behavior of the Fourier transform at large $Q$, as subleading terms get contributions from higher orders in the EFT expansion as well as non-perturbative effects. We therefore obtain the resulting leading order behavior
\begin{equation}
    \langle \mathcal{Q}(\mathbf{n}_1) \mspace{1mu} \mathcal{Q}(\mathbf{n}_2) \rangle \approx \bigg(\frac{Q}{\Omega_{d-2}}\bigg)^2 \Big( 1 + O(1/\Delta_Q) \Big) \qquad (Q \to \infty).
\end{equation}
We can repeat this same procedure for the energy flux two-point function, obtaining
\begin{equation}
    \langle \mathcal{E}(\mathbf{n}_1) \mspace{1mu} \mathcal{E}(\mathbf{n}_2) \rangle \approx \bigg(\frac{E}{\Omega_{d-2}}\bigg)^2 \Big( 1 + O(1/\Delta_Q) \Big) \qquad (Q \to \infty).
\end{equation}

We thus found that in any $U(1)$ symmetric CFT satisfying the broad hypotheses of the large charge expansion,  the lowest-dimension operator $\Ocal_Q$  with  charge $Q$ creates a state of homogeneous charge and energy density as $Q \to \infty$, with any inhomogeneities suppressed by $1/\Delta_Q$.

\subsection{Generalization to Higher-Point Functions}

This analysis can be extended to correlation functions involving an arbitrary number of flux operator insertions, so long as the number of operators $k$ is sufficiently small compared to the total charge $Q$. Starting with the $k$-point function on the cylinder, the leading semiclassical result is simply a product of $k$ expectation values,
\be
\frac{\<Q| \mspace{1mu} J_{\mu_1}(\tau_1,\vec{N}_1) \mspace{1mu} \cdots \mspace{1mu} J_{\mu_k}(\tau_k,\vec{N}_k) \mspace{1mu}|Q\>}{\<Q|Q\>} = \<J_{\mu_1}\> \cdots \<J_{\mu_k}\>,
\ee
with an analogous expression for $k$ insertions of the stress tensor.

Because this $k$-point function factorizes into a product of two- and three-point functions, we can again easily map to the Euclidean plane, Wick rotate, and integrate to obtain a similarly factorized Lorentzian correlator in $x$-coordinates:
\begin{equation}\label{eq:HigherPtPos}
    \frac{\<\overline{\Ocal}_Q(x_f) \mspace{1mu} \mathcal{Q}(\mathbf{n}_1) \mspace{1mu} \cdots \mspace{1mu} \mathcal{Q}(\mathbf{n}_k) \mspace{1mu} \Ocal_Q(x_i)\>}{\<\overline{\Ocal}_Q(x_f) \mspace{1mu} \Ocal_Q(x_i)\>} = \bigg(\frac{Q}{\Omega_{d-2}}\bigg)^k \frac{x_{fi}^{k(d-2)}}{(n_1 \cdot x_{fi})^{d-2} \cdots (n_k \cdot x_{fi})^{d-2}}.
\end{equation}
While the Fourier transform of this general correlator to momentum space is unknown, we can still determine its leading behavior at large $Q$, as explained in detail in appendix~\ref{sec: appendix details computation}. The result is
\begin{equation}
    \langle \mathcal{Q}(\mathbf{n}_1) \mspace{1mu} \cdots \mspace{1mu} \mathcal{Q}(\mathbf{n}_k) \rangle \approx \bigg(\frac{Q}{\Omega_{d-2}}\bigg)^k \Big( 1 + O(k^2/Q^{\frac{d}{d-1}}) \Big) \qquad (Q \to \infty),
\end{equation}
with a similar expression for the energy flux $k$-point function,
\begin{equation}
    \langle \mathcal{E}(\mathbf{n}_1) \mspace{1mu} \cdots \mspace{1mu} \mathcal{E}(\mathbf{n}_k) \rangle \approx \bigg(\frac{E}{\Omega_{d-2}}\bigg)^k \Big( 1 + O(k^2/Q^{\frac{d}{d-1}}) \Big) \qquad (Q \to \infty).
\end{equation}
So long as $k \ll \sqrt{\Delta_Q}$, we therefore expect all flux correlation functions to be homogeneous in the background of the large charge operator $\Ocal_Q$.


\subsection{Large Charge at the Wilson-Fisher Fixed Point}

The EFT calculation presented above describes the universal large charge regime for a broad class of CFTs that meet the generic hypotheses outlined in Section~\ref{semiclassics}. As demonstrated in~\cite{Badel:2019oxl}, this EFT can be effectively applied to the Wilson-Fisher fixed points in the asymptotic regime $\lambda_* n\gg 1$.  On the other hand, Wilson-Fisher models admit a semiclassical description for arbitrary values of $\lambda_* n$ (as long as $n\gg 1$), offering a well-defined framework to thoroughly explore the transition from the {\it{few}} ($\lambda_* n\ll 1$) to the {\it {many}}  ($\lambda_* n\gg 1$) quanta regime.
This subsection is devoted to the simplest Wilson-Fisher model consisting of a complex scalar field $\phi$ with quartic coupling $\frac{\lambda}{4} (\Bar{\phi}\phi)^2$ at its fixed point in $d=4-\epsilon$~\cite{Badel:2019oxl}.\footnote{Other possible weakly coupled models offering a UV completion of the superfluid EFT include the $\mathds{C}P^{N-1}$  model at large N \cite{DeLaFuente:2018uee}, or the supersymmetric fixed point with a single chiral superfield $\Phi$ and superpotential $W = \Phi^3$ \cite{Hellerman:2015nra}.} 
The UV completeness of the Wilson-Fisher model also allows us to bypass the issue of the potential breakdown of the EFT description, which we discuss in Section~\ref{sec: EFT breaking}.

Working directly on the cylinder (we again take $R=1$), the action for the $O(2)$ Wilson-Fisher model reads:
\begin{equation}
    S= \int d^dx \sqrt{g}\left[g^{\mu\nu}\partial_\mu \Bar{\phi}\partial_\nu\phi + \bigg(\frac{d-2}{2}\bigg)^2 \mspace{1mu} \Bar{\phi}\phi +\frac{\lambda_*}{4} \mspace{1mu} (\Bar{\phi}\phi)^2 \right],
\end{equation}
where the mass term is dictated by conformal invariance and controlled by the spatial curvature of the cylinder.
It is convenient to reexpress the complex field as
\begin{equation}
    \phi =\frac{\rho}{\sqrt{2}}e^{i\chi}.
\end{equation}
In the charge $n$ ground state created by $\phi^n$, the spatially homogeneous solution of the equations of motion takes the form
\begin{equation}\label{eqn: solution on cylinderWF}
    \rho = f = \textnormal{constant},  \qquad \chi = -i\mu \tau + \textnormal{constant},
\end{equation}
with the parameters satisfying the constraints
 \begin{equation}
\mu^2- \left(\frac{d-2}{2}\right)^2=\frac{\lambda_*}{4}f^2, \qquad \mu f^2 \Omega_{d-1} = n.
 \end{equation}
 The first constraint corresponds to the equations of motion, while the second fixes the total charge to $n$.

Expanding the action at quadratic order in the two independent fluctuations $\sigma(x)$ and $\pi(x)$
\begin{equation}\label{expandWF}
    \rho(x)= f+ \sigma(x), \qquad \chi(x) =-i \mu \tau  +\frac{\pi(x)}{\sqrt{2}f},
\end{equation}
we obtain the spectrum
\begin{equation}
    \omega^2_\pm(\ell)= J_\ell^2 + 3\mu^2-\big(\tfrac{d-2}{2}\big)^2 \pm  \sqrt{\Big(3\mu^2-\big(\tfrac{d-2}{2}\big)^2\Big)^2 +4 J_\ell^2 \mu^2},
\end{equation}
with $J_\ell^2 \equiv \ell(\ell+d-2)$. Taking the large $\mu$ limit (or equivalently $\lambda n \gg 1$), the $\omega_+(\ell)$ modes decouple as they exhibit a gap $\mu$, while the $\omega_-(\ell)$ modes reduce  to the spectrum of the Goldstone mode of the EFT.  The EFT Lagrangian \eqref{eq:Leff} is recovered by integrating out the radial mode $r(x)$.

Like  in the EFT, the operators are local functions of the fields. The $U(1)$ current is \begin{equation}
    J_\mu=i \rho^2 \partial_\mu \chi,
\end{equation}
while the stress tensor is
\begin{equation}
    \begin{split}
        T_{\mu \nu}=& \rho^2 \partial_\mu \chi \partial_\nu \chi +\frac{2}{3}\partial_\mu \rho \partial_\nu \rho\ -\frac{1}{3} \rho \partial_\mu\partial_\nu \rho \\
        -& g_{\mu \nu} \bigg[ \frac{\rho^2}{2} (\partial \chi)^2 + \frac{1}{2}(\partial \rho)^2 -\frac{1}{6} \partial_\sigma\partial^\sigma \rho^2 + \frac{1}{2}\bigg(\frac{d-2}{2}\bigg)^2 \rho^2 + \frac{\lambda_*}{16} \rho^4 \bigg].
    \end{split}
\end{equation}
In the limit where the radial mode decouples, we recover eqs.~\eqref{eq:Jeft} and \eqref{eq:Teft}.

The expectation values of $J_\mu$ and $T_{\mu \nu}$ in the background of $\phi^n$, when expressed in terms of the quantum numbers $n$ and $\Delta_n$, coincide with the EFT result in eq.~\eqref{eq: J and T expectation value} for \emph{any} value of $\lambda_* n \simeq \epsilon n$, not only for $\lambda_* n \gg 1$. This is unsurprising as this result is dictated by Ward identities. At leading order in the semiclassical expansion, the higher-point current correlators therefore exactly match those of the EFT found in Section~\ref{sec: flux correlator EFT},  as they are simply given by products of the expectation values. Repeating the same analysis as for the EFT, we therefore find
\begin{equation}
    \hat h^{(0)}(\epsilon n,\theta)= 0.
\end{equation}
Crucially, this result holds for arbitrary  $\epsilon n$ ($\simeq \lambda_* n$), as long as $n \gg 1$.

Besides offering an explicit realization of the generic EFT, the Wilson-Fisher theory enables us to study better the transition from few to many quanta. Contrary to what happens for the scaling dimension $\Delta_Q$ studied in \cite{Badel:2019oxl}, which exhibits qualitatively different behavior in the two regimes, we find that the leading correction to homogeneity $\hat h^{(0)}(\epsilon n,\theta)$ vanishes identically in both regimes.


\subsection{Breakdown of Lorentzian EFT} \label{sec: EFT breaking}

In Section~\ref{sec: flux correlator EFT}, we derived the Lorentzian correlators by first computing on the Euclidean cylinder at leading order in the EFT and then analytically continuing the result to the Lorentzian plane. However, the computation on the Euclidean cylinder is reliable only in the kinematical regime of validity of the EFT. With the charge $Q$ and $-Q$ operators inserted at respectively $\tau\to -\infty$ and $\tau \to +\infty$ (see section \ref{semiclassics}), the EFT regime corresponds to separations between individual insertions of the current or stress tensor that are larger than $1/\mu$. Equivalently, the EFT is defined by an expansion in powers of $\partial/\mu$. In particular, for the insertion of just two currents, as in eq.~\eqref{eq:semiclassicJJcyl}, the  condition takes the form
\be
\label{eftvalidity}
|\tau_1-\tau_2|^2+ \big|\arccos(\vec{N}_1\cdot \vec{N}_2)\big|^2 \gtrsim \frac{1}{\mu^2}.
\ee
Considering the corresponding correlator on the plane
\begin{equation*}
\langle \bar{\cal O}_Q(x_f) \mspace{1mu} J(x_1) \mspace{1mu} J(x_2) \mspace{1mu} {\cal O}_Q(x_i)\rangle,
\end{equation*}
eq.~\eqref{eftvalidity} is phrased in terms of  the conformal ratios
\begin{equation}
u\equiv \frac{x_{1i}^2\mspace{1mu}x_{2f}^2}{x_{1f}^2\mspace{1mu}x_{2i}^2}\,,\qquad v\equiv \frac{x_{12}^2\mspace{1mu}x_{if}^2}{x_{1f}^2\mspace{1mu}x_{2i}^2} \, , \qquad \Big(x_{ab}^2\equiv (x_a-x_b)^2\Big),
\end{equation}
as
\begin{equation}
\frac{\log^2(u)}{4}+\arccos^2\left(\frac{1+u-v}{2\sqrt{u}}\right) \gtrsim \frac{1}{\mu^2}\,.
\end{equation}
For $\mu \gg 1$ the above condition can equivalently, and more simply, be expressed as 
\begin{equation}\label{uandv}
    |1-u| \gtrsim \frac{1}{\mu} \quad\textnormal{and/or}\quad v \gtrsim \frac{1}{\mu^2}\,.
\end{equation}

\begin{figure}[t!]
    \centering
    \includegraphics[scale=0.7]{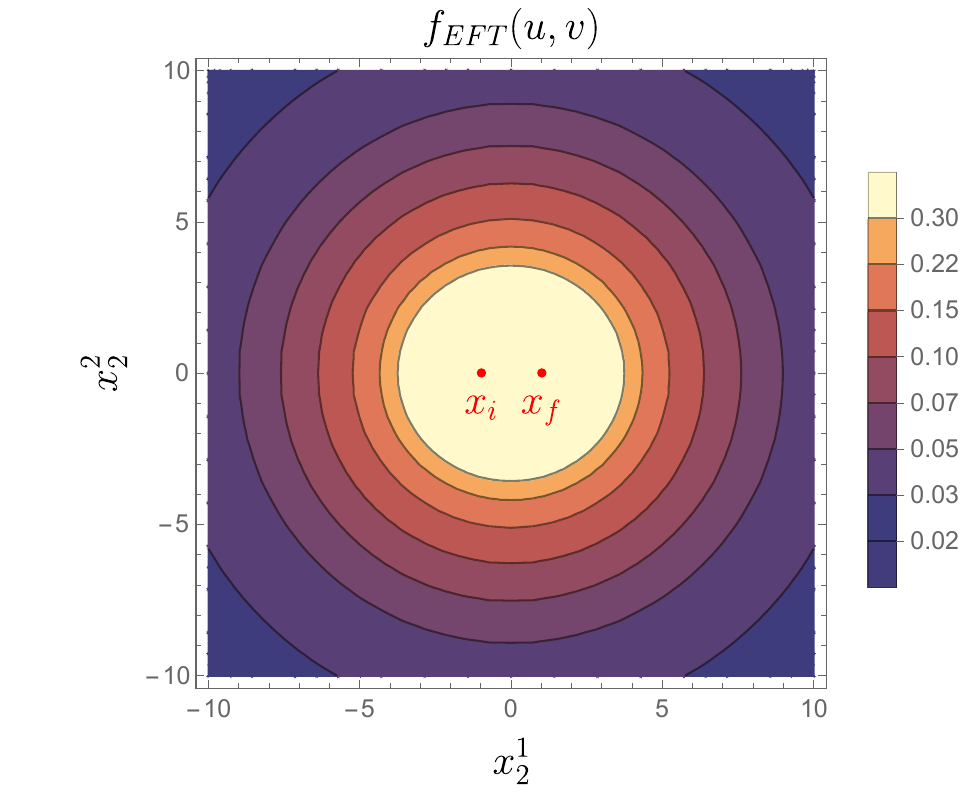}
    \caption{Value of the conformal invariant $f_{EFT}(u,v)\equiv\frac{\ln^2(u)}{4}+\arccos^2\left(\frac{1+u-v}{2\sqrt{u}}\right)$ as a function of the position of the insertion $x_2$ in the Euclidean plane. For simplicity, the function is plotted in two spatial dimensions $x_2\equiv(x_2^1,x_2^2)$. $x_i$ and $x_f$  are respectively chosen to be $(-1,0)$, $(1,0)$, represented on the plot as the two red points, and $x_1$ is sent to infinity. The EFT is valid for computing the four-point function as long as $f_{EFT}(u,v) \gtrsim \frac{1}{\mu^2}$. We observe that this is the case as long as $x_2$ is approximately inserted in a compact region of radius $\sim \mu |x_{if}|$, close to the ``source'' insertions $x_{i/f}$.  }
    \label{fig:eft breaking euclidean}
\end{figure}

It is instructive to study these conditions for finite Euclidean $x_i$ and $x_f$, a choice similar to the one implied by the form of the state in eq.~\eqref{eqn: state} when working in the Lorentzian plane. Using conformal transformations one can see that the coordinate choices satisfying eq.~\eqref{uandv} can  always essentially be mapped into a configuration where $x_1$ and $x_2$ are located in a compact region of size
$\sim \mu x_{if}$ encircling $x_i$ and $x_f$, while remaining sufficiently separated, with roughly  $x_{12}\gtrsim x_{if}/\mu$. This result is physically compatible with a picture where the insertion of ${\cal O}_Q(x_i)$ and $\bar {\cal O}_Q(x_f)$ creates a medium with a density of charge and energy, which is large in the surrounding region and decays at infinity. The power law nature of this decay is simply dictated by the 3-point function $\langle \bar {\cal O}_Q J {\cal O}_Q\rangle$. According to this picture, it is then physically intuitive that the semiclassical description works in the region of high density $x_{12}\lesssim x_{if}/\mu$, so long as  
$x_{12}$ is larger than the typical length $ x_{if}/\mu$ associated with the medium density. Notice that, simply applying an inversion starting from the above configuration, we can always map one of the two observation points, say $x_1$, to $\infty$. As shown in Figure~\ref{fig:eft breaking euclidean}, the region of EFT validity then simply consists of the points where $x_2$ is in a ball of size $\mu x_{if}$ enclosing $x_i$ and $x_f$.

When considering the analytic continuation to Lorentzian signature, the range of $u$ and $v$ is extended to include all real values, not just positive ones. By analyticity, we expect the domain of validity of the EFT description to still be defined by eq.~\eqref{uandv}, with an added absolute value for $v$.\footnote{Because in Lorentzian signature we also expect oscillatory behavior, this requirement should more precisely be stated in integral form: only quantities smeared over regions of $|1-u|$ and $\sqrt{|v|}$ larger than $1/\mu$ can be reliably computed within the EFT description.}
Now, the flux correlators we have considered in this work are (partially) integrated quantities with respect to the coordinates defining $u$ and $v$.
As we will now discuss, the integration region extends outside the domain defined by eq.~\eqref{uandv}, so that the flux correlators receive a contribution that cannot be systematically computed within the EFT, i.e.\ in terms of its Wilson coefficients. However, the region where
eq.~\eqref{uandv} is violated shrinks to a point in the limit $Q\to \infty$ (for which $\mu \to \infty$). Therefore, provided the correlators are not too singular in this region, we still expect this out-of-EFT contribution to  vanish relative to the leading bulk contribution for $Q\to \infty$. It is under this assumption that the  EFT computation done in the previous sections should be taken. While we think this assumption is justified, as indicated by the examples of UV complete models like Wilson-Fisher, we leave to future study a more detailed investigation of the effects due to this vanishing region of integration. In the remainder of this section, we will limit ourselves to discussing the spacetime description of the constraints in eq.~\eqref{uandv} to offer an intuitive physical picture.

The relevant coordinate configuration is one where $x_i$ and $x_f$ are finite (with separation smaller than the length scale $\xi$ defined in 
\eqref{eqn: state}), while $x_1$ and $x_2$ are sent to infinity in a nearly lightlike direction. In light-cone radial coordinates this means $r_1^+\,, r_2^+\to \infty$ with $r_1^-\,, r_2^-$ fixed. 
In this limit, the conformal ratios take the form
\begin{equation}\label{eq: u and v in mink}
        u= \frac{(r_1^- -n_1 \cdot x_i)(r_2^- - n_2 \cdot x_f)}{(r_1^- - n_1 \cdot x_f)(r_2^- - n_2 \cdot x_i)}\,, \qquad
        v=\frac{x_{fi}^2 \, n_1\cdot n_2}{2(r_1^- - n_1 \cdot x_f)(r_2^- - n_2 \cdot x_i)}\,,
\end{equation}
where $n_i$ are the null vectors on the celestial sphere defined below eq.~\eqref{eqn: two point function free x space}.

\begin{figure}[t!]
    \centering
    \includegraphics[scale=0.7]{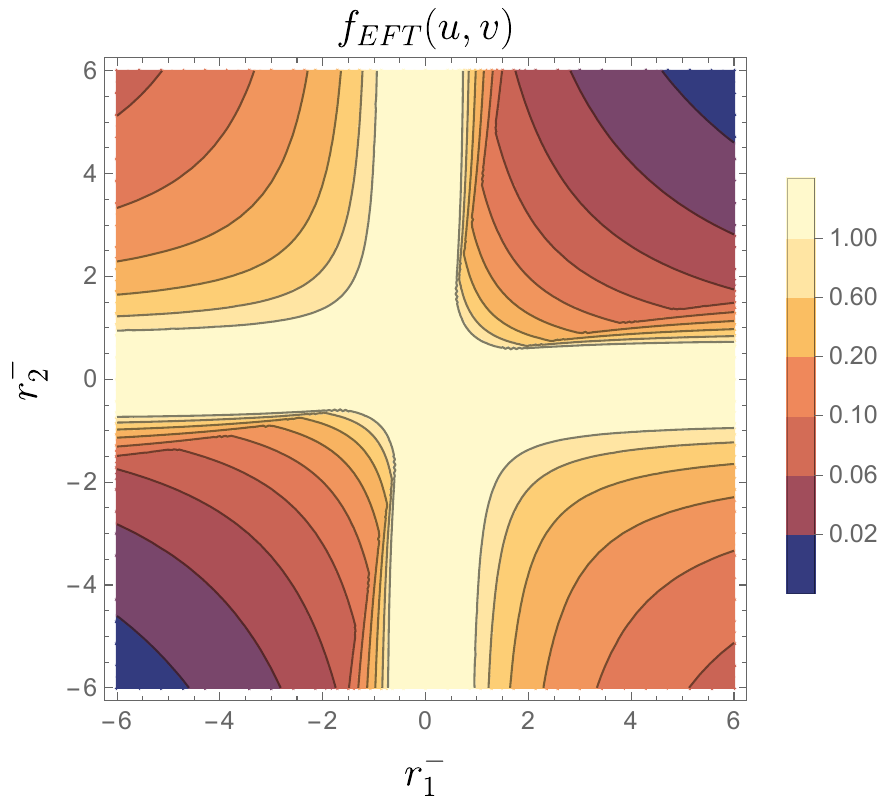}
    \caption{Value of the conformal invariant $f_{EFT}(u,v)\equiv\left|\frac{\ln(u)}{2}\right|^2+\left|\arccos\left(\frac{1+u-v}{2\sqrt{u}}\right)\right|^2$ as a function of $r_1^-$ and $r_2^-$ in the Lorentzian plane.  $x_i=(0,0.5,0,0)$ and $x_f=(0,-0.5,0,0)$ are chosen to be space-like separated, though the result is qualitatively similar if the two points are time-like separated. We chose $n_1=(1,1,0,0)$ and $n_2=(1,0,1,0)$, hence the asymmetry between $r_1^-$ and $r_2^-$. Other choices for $n_{1,2}$ give qualitatively similar results, with the one shown here being the most asymmetric configuration.  The EFT is valid for computing the four-point function as long as $f_{EFT}(u,v)\gtrsim \frac{1}{\mu^2}$.}
    \label{fig:eft breaking minkowski}
\end{figure}

From~\eqref{eq: u and v in mink}, for $x_i\not = x_f$ and $n_1\not = n_2$  one finds that $u\to 1$ and $v\to 0$ as $|r_{1,2}^-|\to \infty$. However, the EFT requirement~\eqref{uandv} is satisfied in a region of the $(r_{1}^-,r_2^-)$ plane that grows with $\mu$. In Figure~\ref{fig:eft breaking minkowski}, we show the shape of that region for spatially separated $x_i$ and $x_f$.\footnote{Notice that there is no correspondence between the Euclidean variables of Figure~\ref{fig:eft breaking euclidean} and the light cone variables of Figure~\ref{fig:eft breaking minkowski}.} One sees that the EFT region corresponds to a choice of $(r_{1}^-,r_2^-)$ such that at least one of the two observation points is near the light cone of the midpoint $(x_i+x_f)/2$.

Like in the Euclidean case, this result has an intuitive physical interpretation. The insertion of the two charged operators essentially creates a shell of width controlled by $x_{fi}$ and centered around $(x_i+x_f)/2$ that expands at the speed of light. One then expects the semiclassical description to apply well for $x_1$ and $x_2$ around the peak of this shell, where the charge density is large, and less so in the tails, where the density goes to zero and non-universal quantum fluctuations may become important. Notice that the EFT situation where both $x_1$ and $x_2$ are around the peak is conformally equivalent to one where only one is (similarly to what happens in the Euclidean case depicted in Figure~\ref{fig:eft breaking euclidean}). This explains the shape of the EFT regions in Figure~\ref{fig:eft breaking minkowski}.

The computation of flux correlators also requires integrating over the coordinates $x_i$ and $x_f$ to comply with the definition~\eqref{eqn: state} of our chosen external state. We expect this integral to be dominated by the region $|x_{fi}| \sim 1/E$, $|x_i+x_f|\lesssim \xi$. The EFT will therefore apply for all configurations except those where both observation points $x_1$ and $x_2$ lie \textit{far outside of a thick light-cone-shaped region} centered at $(x_i +x_f)/2$ and whose thickness is controlled by $|x_{fi}| \sim 1/E$.

As we are dealing with a CFT where signals asymptotically propagate at the speed of light, we expect the bulk of the contribution to the integrated flux to come precisely from the region centered around this thick light cone where the EFT applies, with the non-EFT tail giving a contribution that vanishes $\mu\to \infty$. We have explicitly checked that our computations in free field theory and in the leading semiclassical approximation agree with this expectation for the distribution of flux and are therefore confident that our leading order result from the EFT is physically meaningful. A precise determination of the scaling with $1/Q$ for the leading non-EFT contribution from the tail is instead a more subtle issue that deserves a dedicated study~\cite{workinprogress}.


\section{Summary and Outlook}\label{sec: discussion}

The goal of this work has been to formulate a general procedure for applying the recently developed semiclassical approach for large charge CFT operators to Lorentzian observables. We focused specifically on correlation functions of the charge and energy flux created by spinless large charge operators. The results we obtained cover two related, but different, situations:
\begin{itemize}
\item The leading asymptotic contribution in the inverse  charge expansion of generic $d$-dimensional CFTs possessing a $U(1)$ charge (in this case we indicated the charge by $Q$),
\item  The  limit  $\epsilon \to 0$ with $\epsilon n=$ fixed of Wilson-Fisher fixed points (in this case we indicated the charge by $n$, in keeping with the standard notation for the charged operator, $\phi^n$).
\end{itemize}
The result, expressed by eq.~\eqref{eq:h0result}, is that in these limiting situations the flux correlators are perfectly homogeneous. Also, in view of the free field theory result in section \ref{sec: free theory}, 
 this leading order behavior is physically intuitive, but not trivial. This can be appreciated by approaching the same computation in the Wilson-Fisher fixed point using the standard perturbative Feynman diagram approach. Even in the regime $\epsilon n\ll 1$ the possibility of perturbatively expanding in this parameter can only be inferred by detailed and non-trivial diagrammatics. The basic difficulty, already pointed out in~\cite{Badel:2019oxl}, lies in proving that the contribution of the class of diagrams involving powers of $\epsilon n^2$ (which is in principle $\gg 1$)  exponentiates. The semiclassical approach bypasses these complications and, moreover, provides the answer also in the regime $\epsilon n\gg 1$ where standard perturbation theory irremediably breaks down.

The next step would now be to compute the flux correlator at the next order, which we parametrized in the general case and in the Wilson-Fisher model by respectively $h^{(1)}(d,\theta)$ in eq.~\eqref{eq:Hexp} and $ \hat h^{(1)}(\epsilon n,\theta)$ in eq.~\eqref{eq:hWilsonFisher}. The expectation is that these two functions will have a non-trivial angular dependence thus providing the leading contribution to inhomogeneity. These corrections are simply associated with quantum fluctuations around the semiclassical solution.

\begin{figure}[t!]
    \centering
    \includegraphics[width=0.8\textwidth]{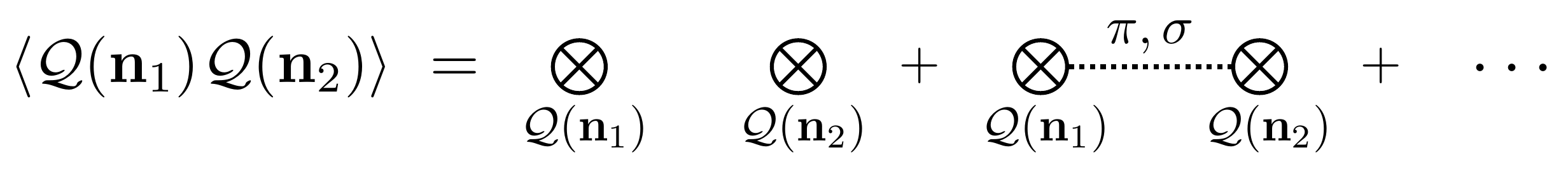}
    \caption{Schematic representation of the semiclassical contributions to the charge flux two-point function. The leading contribution, computed in this work, comes from the product of expectation values $\<\mathcal{Q}\>^2$, while the leading correction comes from the exchange of the fluctuations $\pi$ or $\sigma$. There are also tadpole diagrams which we do not show.}
    \label{fig:CorrectionsQQ}
\end{figure}

In the Wilson-Fisher model, corresponding to the expansion in eq.~\eqref{expandWF},  the current (and similarly $T_{\mu\nu}$) is expanded as
\begin{equation}\label{current_expand}
\begin{aligned}
J_\mu &= f^2\mu \delta_{\mu 0}+2f\left [\sigma\delta_{\mu 0}+ i\frac{1}{2\sqrt 2} \partial_\mu \pi\right ]+\dots\\
&= \frac{n}{\Omega_{d-1}}\left (1+2\sqrt{\frac{\Omega_{d-1}}{n\mu}}\left [\sigma\delta_{\mu 0}+ i\frac{1}{2\sqrt 2} \partial_\mu \pi\right ]+\dots\right )\,.
\end{aligned}
\end{equation}
The quantum fluctuations of $\sigma$ and $\pi$ then lead to a  diagrammatic expansion of the two-point flux correlators, as shown in Figure~\ref{fig:CorrectionsQQ}. The subleading correction is simply controlled by the propagator of the bi-field $(\sigma, \pi)$, while the tadpole diagrams have no effect because the current is not renormalized.
Notice that, by the above equation, one can immediately infer that the correction to the flux correlator has relative size $\Omega_{d-1}/{n\mu}\to 4\pi^2/n\mu$. Since, for $\epsilon n\gg 1$, one has $\mu \sim (\epsilon n)^{1/3}$, we can infer that for $n\to \infty$ the subleading correction scales as $\sim n^{-4/3}$ (this also corresponds to $\lim_{\epsilon n\to \infty} \hat h^{(1)}(\theta,\epsilon n)\sim (\epsilon n )^{-4/3}$).

While the above scaling arguments are straightforward, the computation of the $\theta$ dependence of the correction is not.
The main obstacle is  that we presently do  not possess the propagator in closed form. The propagator can be easily written  as an expansion in the modes over the cylinder. For instance,   the $\langle \pi\pi\rangle$ propagator in the EFT limit $\mu \to\infty$, where $\sigma$ decouples, takes the form
\be
\label{eq:PiPi}
\<\pi(\tau_1,\vec{N}_1) \mspace{1mu} \pi(\tau_2,\vec{N}_2)\> = \frac{1}{c_1d(d-1)\Omega_{d-1}} \bigg( -\frac{1}{2}|\tau_{12}| + \sum_{\ell=1}^\infty \frac{(\ell+\frac{d-2}{2})e^{-\omega_\ell|\tau_{12}|}}{(d-2)\omega_\ell} C_\ell^{(\frac{d-2}{2})}(\vec{N}_1 \cdot \vec{N}_2) \bigg)\,,
\ee
where $C^{(\alpha)}_\ell(x)$ are the Gegenbauer polynomials, and the energies of the individual modes are
\be
\omega_\ell \approx \sqrt{\frac{\ell(\ell+d-2)}{d-1}} + O(1/Q^{\frac{2}{d-1}})\,.
\ee
 On the Euclidean cylinder, this expansion is fast converging for $|\tau_1-\tau_2|\gtrsim 1/\mu$. However, when continuing to Lorentzian signature and integrating over the light-cone variables, the expansion 
does not seem to apply, at least not  straightforwardly. On the other hand, \cite{Giombi:2020enj} offers, for the specific case of the $O(N)$ model at large $N$, an alternative integral representation of the propagator. The adaptation  of the integral form to our case and the study of its possible value in the computation of the flux correlators are certainly tasks worth undertaking.
 
The computation of the next order for the case of a generic $d$-dimensional CFT in the asymptotic large charge regime, which is parametrized by $h^{(1)}(d,\theta)$ in eq.~\eqref{eq:Hexp}, requires instead the tackling of two obstacles. The first, principally technical, is again the computation of the Lorentzian $\langle \pi\pi\rangle$ correlator in a convenient form. The second, more conceptual, concerns the possible contributions that are not captured by the Wilson coefficients of the EFT Lagrangian and which come from the region of integration over the light-cone variables where the EFT breaks down. We gave a detailed illustration of this problem in section \ref{sec: EFT breaking}. There we argued that the contribution from this region should definitely be suppressed as $Q\to \infty$ but we did not attempt an estimate, for which we clearly need additional assumptions. A preliminary study indicates that, under  plausible assumptions on the structure of the UV completion of the EFT,  the contribution from this region is subdominant to that coming from $\langle \pi\pi\rangle$ in the regime of EFT validity. The latter contribution scales as $Q\to \infty$ like $Q^{-d/(d-1)}$, so that in eq.~\eqref{eq:Hexp} we indeed expect $\alpha=\frac{d}{d-1}$. The Wilson-Fisher model offers a specific UV complete EFT where to test this expectation. Indeed, as argued below eq.~\eqref{current_expand}, the next order correction in the model around $d=4$ does scale like $n^{-4/3}$, consistent with the expectation of our preliminary study. 

We consider the facts and the questions reported above as strong motivation for further investigations~\cite{workinprogress}.

\section*{Acknowledgments}    

We are grateful to Nicola Dondi, Bianka Me\c{c}aj, Ian Moult, Filippo Nardi, Domenico Orlando, Yuan Xin, and Sasha Zhiboedov for valuable discussions. 
EF and RR are partially supported by the Swiss National Science Foundation under contract 200020-213104 and through the National Center of Competence in
Research SwissMAP.
MW is supported by the Royal Society-Science Foundation Ireland University Research Fellowship URF{\textbackslash}R1{\textbackslash}221905. AM is supported by the National Science Foundation under Award No.~2310243.


\appendix

\section{Fluxes in Different Frames \label{app:FluxesRelations}}

\begin{figure}[t!]
\bc
\includegraphics[width=0.7\textwidth]{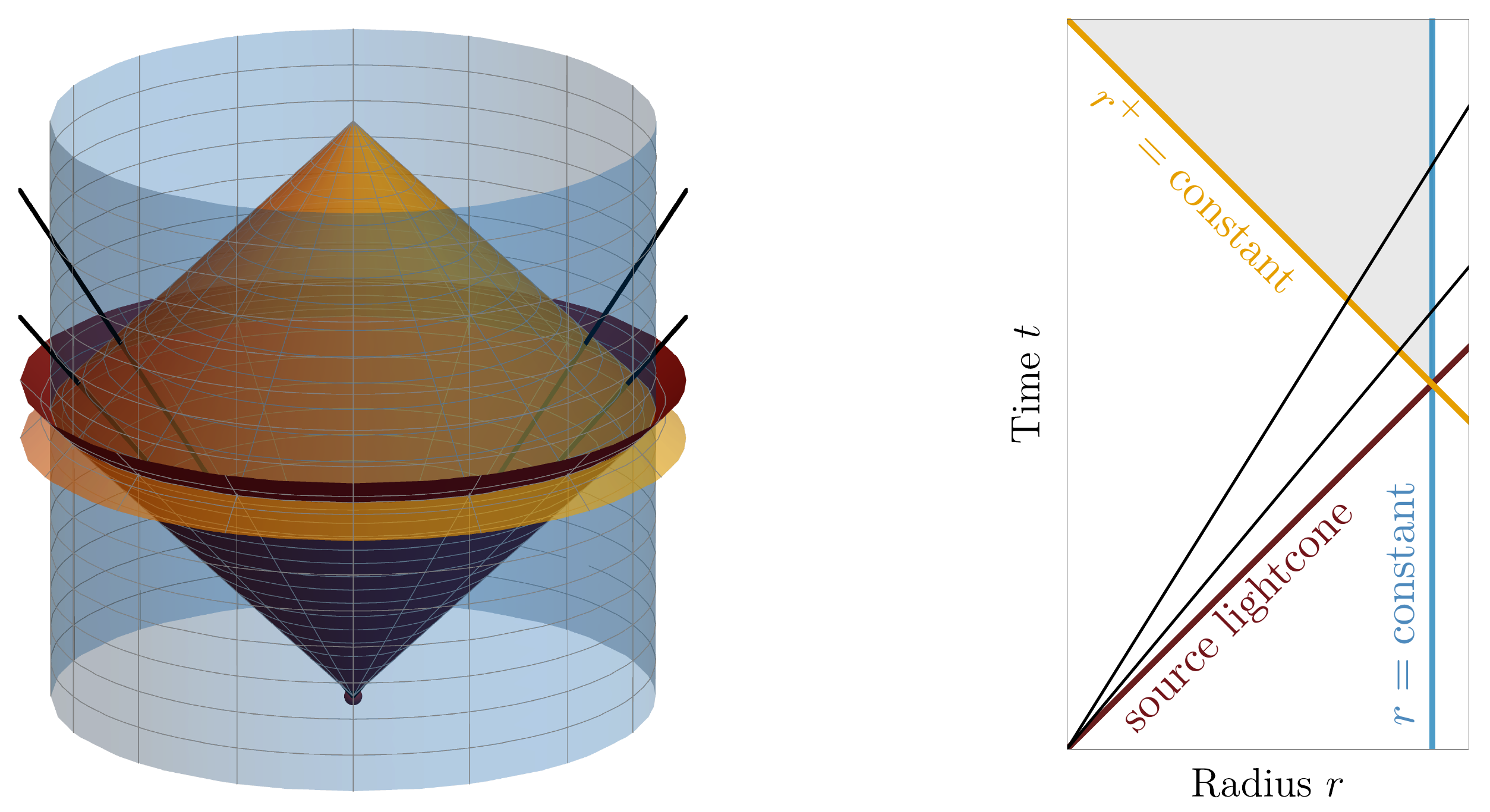}
\ec
\caption{\label{fig:Detectors} \textit{Left:} Different choices for the surface $\Sigma$, where the blue surface corresponds to $r=\textrm{constant}$ while the gold surface corresponds to $r^+=\textrm{constant}$. The red surface is the light cone corresponding to the perturbation. \textit{Right:} Cross-section of this configuration for fixed $\mathbf{n}$. Signals from the perturbation (black lines) eventually cross both surfaces: $r=\textrm{constant}$ (blue) and $r^+=\textrm{constant}$ (gold). Eq.~\eqref{eq:equalfluxes} shows equality of the fluxes integrated over either the gold or blue line bounding the gray shaded region.}
\end{figure}

We would here like to review the basic properties of fluxes and conserved charges that are relevant in our context. Let us consider for that purpose a generic conserved current
\be
\p_\m J^\m = 0\, ,
\label{eq:CurrentConservation}
\ee
which may equally well  be associated to  an internal symmetry or to a spacetime one. In the latter case, we would have $J^\mu=\xi_\nu T^{\mu\nu}$ with $\xi^\nu$ a Killing vector. The conserved charge 
is defined (for a general coordinate choice and with an obvious notation) by
\be
Q= \f{1}{(d-1)!} \int\limits_{\Sigma} dx^{\m_1} \! \wedge \dots \wedge dx^{\m_{d-1}} \sqrt{g} \, \eps_{\m_1 \dots \m_{d-1} \mu}  J^\m(x),
\label{eq:GeneralDefinitionCharge}
\ee
where the integration runs along a $(d-1)$-dimensional hypersurface $\Sigma$. The standard choice for $\Sigma$ is a spacelike surface, typically at constant $t$
\be
Q= \hspace{-1mm} \int\limits_{t=\const} \hspace{-3mm} d\mathbf{x} \, J^0(x)\,.
\label{eq:charge0}
\ee
Current conservation guarantees that $J$ does not depend on the choice of $\Sigma$, provided there is no leakage  at infinity. In suitable physical situations  the charge $J$ may then even be equivalently computed  by choosing  a timelike or null $\Sigma$. Our case, where an originally localized perturbation  eventually spreads (at basically the speed of light)   to infinity, indeed satisfies  that property.
In particular, we can choose $\Sigma$ to be a surface of  constant radius $r$. Indeed, integrating eq.~\eqref{eq:CurrentConservation} over a ball of radius $r$ between times $t_1$ and $t_2$ results in
\be
Q_r(t_1)-Q_r(t_2) = r^{d-2}\int _{t_1} ^ {t_2} dt \int\! d\Omega_{d-2} \, n^i J^i(t,r\mathbf{n})\,,
\ee
where $J_r(t)$ is the charge within the ball at a given time $t$ and $\mathbf{n}$ is a unit vector normal to the surface of the ball.  The spreading of the state to infinity implies $J_r(\infty)=0$ so that 
\be
Q \equiv \lim_{r\to\infty} Q_r(0) = \lim_{r\to\infty}r^{d-2}\int_0^{\infty} dt \int\!d\Omega_{d-2} \, n^i J^i(t,r\mathbf{n})\,.
\label{eq:QtOmega}
\ee
The surface $\Sigma_1$ over which the charge is computed, depicted as the blue cylinder  in Fig.~\eqref{fig:Detectors}, approaches for $r\to\infty$ future null infinity $\mathscr{I}^+$ (see Fig.~\eqref{fig: setup}). Eq.~\eqref{eq:QtOmega} sets the normalization of the total integrated flux, implying in particular eq.~\eqref{inth0}. 

The radial light cone coordinates introduced in section \ref{sec: change of manifold}, offer another natural definition of $J$. That is by integrating over the surface $\Sigma_2$, defined by $r^+=$ constant and depicted by the golden cone in Fig.~\eqref{fig:Detectors}. In the limit $r^+\to \infty$
eq.~\eqref{eq:GeneralDefinitionCharge} becomes
\be
Q = \lim_{r^+\to \infty} \bigg( \frac{r^+}{2} \bigg)^{d-2} \frac{1}{2} \int_{-\infty}^{+\infty} dr^- \int d\Omega_{d-2} \, J^{r^+}(r^+,r^-,\mathbf{n}).
\label{eq:Qr-Omega}
\ee

Now, the surfaces $\Sigma_1$ and $\Sigma_2$ for respectively $r\to\infty$ and $r^+\to\infty$ approach $\mathscr{I}^+$. While on the plane, this horizon is singular, one can use Weyl invariance to map the charges and current to the Lorentzian cylinder, where $\mathscr{I}^+$ is not singular. The currents are therefore smooth on this surface and all components scale similarly. The scaling of the currents at $r^+ \to \infty$ is hence entirely fixed by the change of coordinate from the cylinder to the plane, and this implies $\lim_{r^+\to \infty}J^{r^-}/J^{r^+}=\lim_{r^+\to \infty}J^\perp/J^{r^+}=0$.\footnote{We thank João Penedones for pointing that out to us. } Moreover, using current conservation, one can obtain the scaling of the dominant term $J^{r^+}$ with $r^+$:  $\lim_{r^+\to \infty}(r^+/2)^{d-2} J^{r^+}=2j(r^-,{\mathbf n})$ with $j$  a function independent of $r^+$. This is confirmed by our computation as well as by the study of the three-point function, which is completely fixed by conformal invariance. Synthesizing: on $\mathscr{I}^+$ only the one component of the current, $J^{r^+}\equiv J_{r_-}$, survives and up to a trivial scaling with 
$r_+$ it is described by a function $j(r^-,{\mathbf n})$.

The above result implies asymptotically
\begin{equation}
r^{d-2} n^i J^i(t, r{\mathbf n})\equiv r^{d-2} J^r(t, r{\mathbf n})=\left (\frac{2r}{2r+r^-}\right )^{d-2}j(t-r,{\mathbf n})\xrightarrow{\lim_{r\to \infty}} j(t-r,{\mathbf n})
\end{equation}
and, consequently
\bea
\mathcal{Q}({\mathbf n})&\equiv& \lim_{r^+\to \infty} \bigg( \frac{r^+}{2} \bigg)^{d-2} \frac{1}{2} \int_{-\infty}^{+\infty}  dr^- \, J^{r^+}(r^+,r^-,\mathbf{n})=
 \lim_{r\to \infty}r^{d-2} \int_{0}^{+\infty}  dt \, n^i J^i(t,r\mathbf{n})\\
 &=&\int_{-\infty}^{+\infty}  dr^- j(r^-,{\mathbf n})\,.
\label{eq:equalfluxes}
\eea
This  is stronger than the equality of the two expressions for the charge in eqs.~\eqref{eq:QtOmega} and \eqref{eq:Qr-Omega}: the time-integrated fluxes at each point on the celestial sphere coincide. All this is consistent with the fact that $\mathscr{I}^+$ is not a singular surface in a conformal field theory and thus its integrated flux is  independent of the way this surface is approached.

We can now consider the computation using $y$ coordinates.
For $t \approx r \to \infty$ (or equivalently, $y^+ \to 0^-$), we have the relation
\be
t = \frac{1}{2} \bigg( \f{1+|\vec{y}^\perp|^2}{-y^+} + y^- \bigg), \quad r = \frac{1}{2} \bigg( \f{1+|\vec{y}^\perp|^2}{-y^+} - \bigg(\frac{1-|\vec{y}^\perp|^2}{1+|\vec{y}^\perp|^2} \bigg) y^- \bigg),
\ee
or equivalently,
\be
r^+ = \f{1+|\vec{y}^\perp|^2}{-y^+} - \bigg(\f{|\vec{y}^\perp|^2}{1+|\vec{y}^\perp|^2}\bigg) y^- \approx \f{1+|\vec{y}^\perp|^2}{-y^+}, \quad r^- = \f{y^-}{1+|\vec{y}^\perp|^2}.
\ee
Under this change of coordinates, our definition for the charge flux, both in terms of $t$ and $r^-$ becomes\footnote{Note that when the conserved current corresponds to a higher-spin operator, such as the stress-energy tensor, the behavior of $J_\mu$ under the conformal transformation~\REF{eqn: y versus x coordinates} can involve additional multiplicative factors depending on $\vec{y}^\perp$ (e.g., eq.~\eqref{eq:EnergyExample}).}
\be
    \begin{aligned}
        \mathcal{Q}(\mathbf{n}) &= \lim_{y^+ \to 0^-} \bigg( \f{1+|\vec{y}^\perp|^2}{-2y^+} \bigg)^{d-2} \int_{-\infty}^\infty \frac{dy^-}{1+|\vec{y}^\perp|^2} \cdot (-y^+)^{d-2} (1+|\vec{y}^\perp|^2) J_-(y^+,y^-,\vec{y}^\perp) \\
        &= \bigg( \f{1+|\vec{y}^\perp|^2}{2} \bigg)^{d-2} \int_{-\infty}^\infty dy^- \, J_-(y^+ = 0,y^-,\vec{y}^\perp),
    \end{aligned}
\ee

Taking into account the relation
\be
\vec{n}^\perp = \bigg(\frac{2}{1+|\vec{y}^\perp|^2}\bigg) \vec{y}^\perp,
\ee
we can express the integral over the celestial sphere in terms of $\vec{y}^\perp$ as
\be
d\Omega_{d-2} = \bigg( \f{2}{1+|\vec{y}^\perp|^2} \bigg)^{d-2} d\vec{y}^\perp,
\ee
which guarantees the normalization
\be
Q = \int dy^- d\vec{y}^\perp \mspace{1mu} J_-(y^+=0,y^-,\vec{y}^\perp).
\ee
For completeness, we show how the mapping works for the energy, in which case we have
\be\label{eq:EnergyExample}
E =\int \lefteqn{\overbrace{\phantom{d\Omega_{d-2} \, \bigg( \f{1+|\vec{y}^\perp|^2}{2} \bigg)^{d-2}}}^{d\vec{y}^\perp}}d\Omega_{d-2}
\underbrace{\bigg( \f{1+|\vec{y}^\perp|^2}{2} \bigg)^{d-2}\, \bigg( \f{1+|\vec{y}^\perp|^2}{2} \bigg) \, \overbrace{2 \int dy^- \, T_{--}(y^+=0,y^-,y^\perp)}^{\mc E(\vec{y}^\perp)}}_{\mc E(\mathbf n)}.
\ee

\section{Details of Flux Correlator Computation}\label{sec: appendix details computation}

In this appendix, we present various details of the calculation of flux operator correlators, both for the free scalar and the semiclassical analysis for more general CFTs.

\subsection{Free Theory Calculation in Position Space}

As discussed in section~\ref{sec: free theory}, there are two contributions to both the charge flux two-point function $\<\Qcal\Qcal\>$ and the energy flux two-point function $\<\Ecal\Ecal\>$, corresponding to diagrams $(a)$ and $(b)$ in Figure~\ref{fig:diagram free case n fields}. The $U(1)$ current correlation function corresponding to diagram $(a)$ is
\begin{equation}\label{eq:freeJJA}
    \begin{aligned}
        &\<\bar\phi^n(y_f) \mspace{1mu} J_-(y_1) \mspace{1mu}  J_-(y_2) \mspace{1mu} \phi^n(y_i)\>^{(a)} \\
        & \quad = n^2(n-1)^2 \mspace{2mu} \<\bar\phi(y_f) \mspace{1mu} J_-(y_1) \mspace{1mu} \phi(y_i)\> \mspace{2mu} \<\bar\phi(y_f) \mspace{1mu} J_-(y_2) \mspace{1mu} \phi(y_i)\> \mspace{2mu} \<\bar\phi^{n-2}(y_f) \mspace{1mu} \phi^{n-2}(y_i)\> \\
        & \quad = (-1)^{1-d-n(\frac{d-2}{2})} \frac{n!}{\big((d-2)\Omega_{d-1}\big)^n} \frac{n(n-1)}{(2\Omega_{d-1})^2} \frac{y_{fi}^{(2-n)(d-2)}}{y_{f1}^{d-2} y_{1i}^{d-2} y_{f2}^{d-2} y_{2i}^{d-2}} \bigg( \frac{y_{f1}^+}{y_{f1}^2} + \frac{ y_{1i}^+} {y_{1i}^2}\bigg)  \bigg( \frac{y_{f2}^+} {y_{f2}^2} + \frac{ y_{2i}^+} {y_{2i}^2} \bigg),
    \end{aligned}
\end{equation}
while the current correlation function for diagram (b) is
\begin{equation}\label{eq:freeJJB}
    \begin{aligned}
        &\<\bar\phi^n(y_f) \mspace{1mu} J_-(y_1) \mspace{1mu} J_-(y_2) \mspace{1mu} \phi^n(y_i)\>^{(b)} \\
        & \qquad = n^2 \mspace{2mu} \<\bar\phi(y_f) \mspace{1mu} J_-(y_1) J_-(y_2) \mspace{1mu} \phi(y_i)\> \mspace{2mu} \<\bar\phi^{n-1}(y_f) \mspace{1mu} \phi^{n-1}(y_i)\> \\
        & \qquad = (-1)^{1-d-n(\frac{d-2}{2})} \frac{n!}{\big((d-2)\Omega_{d-1}\big)^n} \frac{n}{(2\Omega_{d-1})^2} \frac{y_{fi}^{(1-n)(d-2)}}{y_{f1}^{d-2} y_{12}^{d-2} y_{2i}^{d-2}} \\
        & \qquad \qquad \times \Bigg( \bigg( \frac{y_{f1}^+} {y_{f1}^2} + \frac{ y_{12}^+} {y_{12}^2} \bigg)  \bigg( \frac{y_{12}^+} {y_{12}^2} + \frac{ y_{2i}^+} {y_{2i}^2} \bigg) + \bigg(\frac{2}{d-2}\bigg) \frac{ (y_{12}^+)^2}{y_{12}^4} \Bigg).
    \end{aligned}
\end{equation}

To obtain the resulting charge flux correlator, we need to integrate over $y_1^-$ and $y_2^-$ for both of these expressions. Starting with diagram $(a)$, we see that eq.~\eqref{eq:freeJJA} factorizes into independent functions of $y_1$ and $y_2$, such that we have
\begin{equation}
    \begin{aligned}
        &\<\bar\phi^n(y_f) \mspace{1mu} \Qcal(\vec{y}_1^\perp) \mspace{1mu} \Qcal(\vec{y}_2^\perp) \mspace{1mu} \phi^n(y_i)\>^{(a)} \\
        & \quad = n^2(n-1)^2 \mspace{2mu} \<\bar\phi(y_f) \mspace{1mu} \Qcal(\vec{y}_1^\perp) \mspace{1mu} \phi(y_i)\> \mspace{2mu} \<\bar\phi(y_f) \mspace{1mu} \Qcal(\vec{y}_2^\perp) \mspace{1mu} \phi(y_i)\> \mspace{2mu} \<\bar\phi^{n-2}(y_f) \mspace{1mu} \phi^{n-2}(y_i)\>.
    \end{aligned}
\end{equation}
We therefore just need to evaluate the integral
\begin{equation}
    \<\bar\phi^n(y_f) \mspace{1mu} \Qcal(\vec{y}_j^\perp) \mspace{1mu} \phi^n(y_i)\> = \frac{(-1)^{\frac{3}{2}-d}}{2(d-2)\Omega^2_{d-1}} \int dy_j^- \frac{1}{y_{fj}^{d-2} y_{ji}^{d-2}} \bigg( \frac{y_{fj}^+}{y_{fj}^2} + \frac{ y_{ji}^+} {y_{ji}^2}\bigg).
\end{equation}
The integrand has two poles, corresponding to the points along the integral where $y_j$ crosses the light cone of $y_i$ or $y_f$. Due to the Wightman ordering of this correlator, the pole associated with $y_i$ is in the upper half-plane while the pole associated with $y_f$ is in the lower half-plane. We can therefore close the integration contour in either direction and obtain the resulting expression
\be
\<\bar\phi(y_f) \mspace{1mu} \Qcal(\vec{y}_j^\perp) \mspace{1mu} \phi(y_i)\> = \frac{(-1)^{2-d} \mspace{1mu} 2^{d-2}}{(d-2)\Omega_{d-1} \Omega_{d-2}} \frac{1}{(y_{fj}^+)^{\frac{d-2}{2}} \mspace{1mu} (y_{ji}^+)^{\frac{d-2}{2}} \mspace{1mu} \Big(y_{fi}^- - \frac{|\vec{y}_{fj}^\perp|^2}{y_{fj}^+} - \frac{|\vec{y}_{ji}^\perp|^2}{y_{ji}^+}\Big)^{d-2}},
\ee
which we can map to $x$-coordinates with the conformal transformation from eq.~\eqref{eqn: y versus x coordinates},
\be
\<\bar\phi(x_f) \mspace{1mu} \Qcal(\mathbf{n}_j) \mspace{1mu} \phi(x_i)\> = \frac{(-1)^{\frac{2-d}{2}}}{(d-2)\Omega_{d-1}\Omega_{d-2}} \frac{1}{(n_j\cdot x_{fi})^{d-2}}.
\ee
We therefore obtain the full expression for diagram $(a)$,
\begin{equation}
    \begin{aligned}
        &\<\bar\phi^n(x_f) \mspace{1mu} \Qcal(\mathbf{n}_1) \mspace{1mu} \Qcal(\mathbf{n}_2) \mspace{1mu} \phi^n(x_i)\>^{(a)} \\
        & \qquad = (-1)^{n(\frac{2-d}{2})} \frac{n!}{\big((d-2)\Omega_{d-1}\big)^n} \frac{n(n-1)}{\Omega_{d-2}^2} \frac{x_{fi}^{(2-n)(d-2)}}{(n_1\cdot x_{fi})^{d-2} \mspace{1mu} (n_2\cdot x_{fi})^{d-2}}.
    \end{aligned}
\end{equation}

Turning to the contribution from diagram $(b)$, we see that eq.~\eqref{eq:freeJJB} does \emph{not} factorize into a product of three-point functions. We instead need to compute
\begin{equation}
    \<\bar\phi^n(y_f) \mspace{1mu} \Qcal(\vec{y}_1^\perp) \mspace{1mu} \Qcal(\vec{y}_2^\perp) \mspace{1mu} \phi^n(y_i)\>^{(b)} = n^2 \mspace{2mu} \<\bar\phi(y_f) \mspace{1mu} \Qcal(\vec{y}_1^\perp) \mspace{1mu} \Qcal(\vec{y}_2^\perp) \mspace{1mu} \phi(y_i)\> \mspace{2mu} \<\bar\phi^{n-1}(y_f) \mspace{1mu} \phi^{n-1}(y_i)\>,
\end{equation}
which contains the double integral
\be
    \begin{aligned}
        & \<\bar\phi(y_f) \mspace{1mu} \Qcal(\vec{y}_1^\perp) \mspace{1mu} \Qcal(\vec{y}_2^\perp) \mspace{1mu} \phi(y_i)\> \\
        & = \frac{(-1)^{2-\frac{3}{2}d}}{4(d-2)\Omega_{d-1}^3} \!\int\! dy_1^- \!\int\! dy_2^- \frac{1}{y_{f1}^{d-2} y_{12}^{d-2} y_{2i}^{d-2}} \Bigg( \! \bigg( \frac{y_{f1}^+} {y_{f1}^2} + \frac{ y_{12}^+} {y_{12}^2} \bigg)  \bigg( \frac{y_{12}^+} {y_{12}^2} + \frac{ y_{2i}^+} {y_{2i}^2} \bigg) + \bigg(\frac{2}{d-2}\bigg) \frac{ (y_{12}^+)^2}{y_{12}^4} \Bigg).
    \end{aligned}
\ee
The simplest way to evaluate these integrals is to close the contour for $y_1^-$ in the lower half-plane (picking up the pole from $y_f$) and close the contour for $y_2^-$ in the upper half-plane (picking up the pole from $y_i$), such that we obtain\footnote{See the appendices of~\cite{Belin:2020lsr} for a detailed discussion of the evaluation of integrals of this form.}
\be
    \begin{aligned}
        & \<\bar\phi(y_f) \mspace{1mu} \Qcal(\vec{y}_1^\perp) \mspace{1mu} \Qcal(\vec{y}_2^\perp) \mspace{1mu} \phi(y_i)\> \\
        & \qquad = \frac{(-1)^{2-d} 2^{d-2}}{(d-2)\Omega_{d-1}\Omega_{d-2}} \frac{1}{(y_{f1}^+)^{\frac{d-2}{2}} \mspace{1mu} (y_{1i}^+)^{\frac{d-2}{2}} \mspace{1mu} \Big(y_{fi}^- - \frac{|\vec{y}_{f1}^\perp|^2}{y_{f1}^+} - \frac{|\vec{y}_{1i}^\perp|^2}{y_{1i}^+}\Big)^{d-2}} \, \delta(\vec{y}_{12}^\perp).
    \end{aligned}
\ee
Using the relation
\be
    \bigg( \frac{1+|\vec{y}^\perp|^2}{2} \bigg)^{2-d} \delta(\vec{y}^\perp) = \delta(\mathbf{n}),
\ee
we can map this expression to $x$-coordinates,
\be
    \<\bar\phi(x_f) \mspace{1mu} \Qcal(\mathbf{n}_1) \mspace{1mu} \Qcal(\mathbf{n}_2) \mspace{1mu} \phi(x_i)\> = \frac{(-1)^{\frac{2-d}{2}}}{(d-2)\Omega_{d-1}\Omega_{d-2}} \frac{1}{(n_1\cdot x_{fi})^{d-2}} \, \delta(\mathbf{n}_1 - \mathbf{n}_2).
\ee
We thus find the resulting expression for diagram $(b)$,
\begin{equation}
    \begin{aligned}
        &\<\bar\phi^n(x_f) \mspace{1mu} \Qcal(\mathbf{n}_1) \mspace{1mu} \Qcal(\mathbf{n}_2) \mspace{1mu} \phi^n(x_i)\>^{(b)} \\
        & \qquad = (-1)^{n(\frac{2-d}{2})} \frac{n!}{\big((d-2)\Omega_{d-1}\big)^n} \frac{n}{\Omega_{d-2}} \frac{x_{fi}^{(1-n)(d-2)}}{(n_1\cdot x_{fi})^{d-2}} \, \delta(\mathbf{n}_1 - \mathbf{n}_2).
    \end{aligned}
\end{equation}
This particular diagram therefore only contributes if $\mathbf{n}_1 = \mathbf{n}_2$. Intuitively, this is because both insertions of $\Qcal$ are ``measuring'' the same free particle, which can only occur if they're located at the same point on the celestial sphere.

This exact same procedure can be repeated for the energy flux correlator, resulting in the contribution
\begin{equation}
    \begin{aligned}
        &\<\bar\phi^n(x_f) \mspace{1mu} \Ecal(\mathbf{n}_1) \mspace{1mu} \Ecal(\mathbf{n}_2) \mspace{1mu} \phi^n(x_i)\>^{(a)} \\
        & \qquad = (-1)^{n(\frac{2-d}{2})-1} \frac{n!}{\big((d-2)\Omega_{d-1}\big)^n} \frac{n(n-1)(d-2)^2}{\Omega_{d-2}^2} \frac{x_{fi}^{(2-n)(d-2)}}{(n_1\cdot x_{fi})^{d-1} \mspace{1mu} (n_2\cdot x_{fi})^{d-1}},
    \end{aligned}
\end{equation}
from diagram $(a)$ and
\begin{equation}
    \begin{aligned}
        &\<\bar\phi^n(x_f) \mspace{1mu} \Ecal(\mathbf{n}_1) \mspace{1mu} \Ecal(\mathbf{n}_2) \mspace{1mu} \phi^n(x_i)\>^{(b)} \\
        & \qquad = (-1)^{n(\frac{2-d}{2})-1} \frac{n!}{\big((d-2)\Omega_{d-1}\big)^n} \frac{n(d-2)^2}{\Omega_{d-2}} \frac{x_{fi}^{(1-n)(d-2)}}{(n_1\cdot x_{fi})^{d-1}} \mspace{1mu} \delta(\mathbf{n}_1 - \mathbf{n}_2),
    \end{aligned}
\end{equation}
from diagram $(b)$.

\subsection{Fourier Transform to Momentum Space}

Once we've obtained a flux operator correlation function in position space, we need to Fourier transform with respect to the positions $x_i$ and $x_f$ of the external operators to obtain the desired momentum space expectation value. Based on the structure of the $k$-point function in eq.~\eqref{eq:HigherPtPos}, we must evaluate the general Fourier transform
\be
    F_k^{(\alpha,\beta)}(\mathbf{n}_1,\ldots,\mathbf{n}_k;p) \equiv \int d^dx \, e^{i p \cdot x} \frac{1}{x^{2\alpha} \mspace{1mu} (n_1 \cdot x)^\beta \mspace{1mu} \cdots \mspace{1mu} (n_k \cdot x)^\beta}.
\ee
Our strategy will be to evaluate this expression recursively with the convolution theorem. We therefore need to first evaluate the Fourier transforms of the individual terms
\be
    f_\alpha(p) \equiv \int d^dx \, \frac{e^{ip\cdot x}}{\big((x^+ - i\epsilon)(x^- - i\epsilon) - |\vec{x}^\perp|^2\big)^\alpha} = \frac{(-1)^\alpha \pi^{\frac{d+2}{2}} p^{2\alpha-d}}{2^{2\alpha-d-1} \Gamma(\alpha) \Gamma(\alpha-\frac{d-2}{2})} \Theta(p_0) \Theta(p^2),
\ee
and
\be
    g_\beta(\mathbf{n};p) \equiv \int d^dx \, \frac{e^{ip\cdot x}}{(n \cdot x)^\beta} = \frac{(2\pi)^d}{\Gamma(\beta)} p_0^{\beta-1} \delta(\mathbf{p} - p_0 \mathbf{n}) \, \Theta(p_0),
\ee
where $\Theta(x)$ is the Heaviside step function.

Using these building blocks, we can evaluate the Fourier transform for $k=1$,
\be
    \begin{aligned}
        F_1^{(\alpha,\beta)}(\mathbf{n};p) &= \int \frac{d^dq}{(2\pi)^d} \, f_\alpha(p-q) \, g_\beta(\mathbf{n};q) \\
        &= \frac{1}{\Gamma(\beta)} \int_0^{\frac{p^2}{2n\cdot p}} \!dq_0 \, q_0^{\beta-1} f_\alpha(p-q_0 n) \\
        &= \frac{(-1)^\alpha \pi^{\frac{d+2}{2}}}{2^{2\alpha+\beta-d-1} \Gamma(\alpha)\Gamma(\alpha+\beta-\frac{d-2}{2})} \frac{p^{2(\alpha+\beta)-d}}{(n \cdot p)^\beta} \Theta(p_0) \Theta(p^2),
    \end{aligned}
\ee
where we have used the fact that $f_\alpha(p-q) \propto \Theta\big((p-q)^2\big)$ to limit the integral over $q_0$ to a finite range.

We can then apply the convolution theorem again to obtain the Fourier transform of $k=2$ for the simplified case where $p^\mu=(E,\mathbf{0})$
\be
    \begin{aligned}
        &F_2^{(\alpha,\beta)}(\mathbf{n}_1,\mathbf{n}_2;E) = \frac{1}{\Gamma(\beta)} \int_0^{\frac{E}{2}} \!dq_0 \, q_0^{\beta-1} F_1^{(\alpha,\beta)}(\mathbf{n}_1;p-q_0 n_2) \\
        & \qquad =\frac{ (-1)^\alpha \pi^{\frac{d+2}{2}} E^{2\alpha+2\beta-d}}{2^{2\alpha+2\beta-d-1} \Gamma(\alpha) \Gamma(\alpha+2\beta-\frac{d-2}{2})} \, {}_2F_1\big(\beta,\beta;\alpha+2\beta-\tfrac{d-2}{2};\tfrac{n_1 \cdot n_2}{2}\big) \, \Theta(E).
    \end{aligned}
\ee
Unfortunately, the necessary integrals become increasingly complicated for higher $k$. However, the expression simplifies dramatically if we focus on the limit $\alpha \to \infty$ with $\beta$ and $k$ fixed. Looking at the integral for $k=2$ again (for general $p^\mu$), we have
\be
    \begin{aligned}
        &F_2^{(\alpha,\beta)}(\mathbf{n}_1,\mathbf{n}_2;p) \\
        & \qquad = \frac{(-1)^\alpha \pi^{\frac{d+2}{2}}}{2^{2\alpha+\beta-d-1} \Gamma(\alpha)\Gamma(\beta)\Gamma(\alpha+\beta-\frac{d-2}{2})} \int_0^{\frac{p^2}{2n_2\cdot p}} \!dq_0 \, \frac{q_0^{\beta-1}(p^2-2q_0 \mspace{1mu} n_2 \cdot p)^{\alpha+\beta-\frac{d}{2}}}{(n_1 \cdot p - q_0 \mspace{1mu} n_1 \cdot n_2)^\beta}.
    \end{aligned}
\ee
If we interpret the factor of $(p^2-2q_0 \mspace{1mu} n_2 \cdot p)^{\alpha+\beta-\frac{d}{2}}$ as an overall derivative and integrate by parts, the boundary terms both vanish due to the positive powers of $q_0$ and $p^2-2q_0 \mspace{1mu} n_2 \cdot p$, leaving us with the new integral
\be
    \begin{aligned}
        & \int_0^{\frac{p^2}{2n_2\cdot p}} \!dq_0 \, \frac{q_0^{\beta-1}(p^2-2q_0 \mspace{1mu} n_2 \cdot p)^{\alpha+\beta-\frac{d}{2}}}{(n_1 \cdot p - q_0 \mspace{1mu} n_1 \cdot n_2)^\beta} \\
        & \quad = \frac{1}{2 \mspace{1mu} n_2 \cdot p(\alpha+\beta-\frac{d-2}{2})} \int_0^{\frac{p^2}{2n_2\cdot p}} \!dq_0 \, \bigg( (\beta-1) \frac{q_0^{\beta-2}(p^2-2q_0 \mspace{1mu} n_2 \cdot p)^{\alpha+\beta-\frac{d-2}{2}}}{(n_1 \cdot p - q_0 \mspace{1mu} n_1 \cdot n_2)^\beta} \\
        & \hspace{7cm} + \, \beta \mspace{1mu} n_1 \cdot n_2 \mspace{1mu} \frac{q_0^{\beta-1}(p^2-2q_0 \mspace{1mu} n_2 \cdot p)^{\alpha+\beta-\frac{d-2}{2}}}{(n_1 \cdot p - q_0 \mspace{1mu} n_1 \cdot n_2)^{\beta+1}} \bigg).
    \end{aligned}
\ee
As we can see, each integration by parts brings down an overall factor of $1/\alpha$. The leading contribution as $\alpha \to \infty$ will therefore come from the fewest actions of integration by parts that lead to a non-vanishing boundary term, which is obtained by repeatedly acting with derivatives on the initial power of $q_0^{\beta-1}$ in the numerator until its exponent vanishes (as $\beta$ is a positive integer). We thus find the leading behavior at large $\alpha$:
\be
    F_2^{(\alpha,\beta)}(\mathbf{n}_1,\mathbf{n}_2;p) \approx \frac{ (-1)^\alpha \pi^{\frac{d+2}{2}}}{2^{2\alpha+2\beta-d-1} \Gamma(\alpha) \Gamma(\alpha+2\beta-\frac{d-2}{2})} \frac{p^{2\alpha+4\beta-d}}{(n_1 \cdot p)^\beta \mspace{1mu} (n_2 \cdot p)^\beta} \Big( 1 + O(1/\alpha) \Big).
\ee
We can apply this same analysis recursively to higher values of $k$, obtaining the general leading order behavior
\be
    F_k^{(\alpha,\beta)}(\mathbf{n}_1,\ldots,\mathbf{n}_k;p) \approx \frac{ (-1)^\alpha \pi^{\frac{d+2}{2}}}{2^{2\alpha+k\beta-d-1} \Gamma(\alpha) \Gamma(\alpha+k\beta-\frac{d-2}{2})} \frac{p^{2\alpha+2k\beta-d}}{(n_1 \cdot p)^\beta \mspace{1mu} \cdots \mspace{1mu} (n_k \cdot p)^\beta} \quad (\alpha \to \infty).
\ee
The next-to-leading order correction comes from acting with one derivative on any of the factors of $(n_i \cdot p - q_0 \mspace{1mu} n_i \cdot n_j)^{\beta+1}$ in the denominator while integrating by parts and is, therefore, $O(k^2/\alpha)$ relative to the leading behavior (with a suppression of $1/\alpha$ due to the extra derivative and an enhancement of $k(k-1)$ due to combinatorics).

\subsection{Map from the Cylinder to the Plane}

In order to use the semiclassical approach of~\cite{Hellerman:2015nra,Monin:2016jmo} for the calculation of flux operators, including subleading corrections, we need to map expectation values on the Euclidean cylinder to correlation functions on the Euclidean plane. Here we'll present the details of this mapping, focusing mainly on the two-point function of the $U(1)$ current $J_\mu$ before discussing the generalization to higher-point functions and correlators of $T_{\mu\nu}$.

Our starting point is the cylinder four-point function from eq.~\eqref{eq:semiclassicJJcyl}:
\be
\frac{\<Q|\mspace{1mu} J^\mu(\tau_1,\vec{N}_1) \mspace{1mu} J^\nu(\tau_2,\vec{N}_2) \mspace{1mu}|Q\>}{\<Q|Q\>} = - \bigg( \frac{Q}{\Omega_{d-1}} \bigg)^2 \delta^\mu_{\,\tau} \delta^{\nu}_{\,\tau}.
\ee
Because all four operators are primary, we can use the Weyl transformation in eq.~\eqref{eq:WeylTransform} to obtain the correlator on the plane\footnote{For notational simplicity, in this appendix we suppress the ``$E$'' subscript on $y$-coordinates, with the understanding that \emph{all} correlators here are Euclidean.}
\be\label{eq:JJplane}
    \begin{aligned}
        \frac{\<\overline{\Ocal}_Q(\infty) \mspace{1mu} J^\mu(y_1) \mspace{1mu} J^\nu(y_2) \mspace{1mu} \Ocal_Q(0)\>}{\<\overline{\Ocal}_Q(\infty) \mspace{1mu} \Ocal_Q(0)\>} &= e^{-(d-1)(\tau_1 + \tau_2)} \mspace{1mu} \frac{\<Q| \mspace{1mu} J^\mu(\tau_1,\vec{N}_1) \mspace{1mu} J^\nu(\tau_2,\vec{N}_2) \mspace{1mu}|Q\>}{\<Q|Q\>} \\
        &= - \bigg( \frac{Q}{\Omega_{d-1}} \bigg)^2 \frac{y_1^\mu}{y_1^d} \frac{y_2^\nu}{y_2^d}.
    \end{aligned}
\ee

We now need to perform a conformal transformation to move the external operators away from zero and infinity to arbitrary positions $y_i$ and $y_f$. The most systematic way to do this is to decompose this correlation function in terms of the basis of tensor structures allowed by conformal symmetry. For a four-point function with two scalar operators and two insertions of the $U(1)$ current, there are five independent structures, allowing us to write any such correlator in the general form~\cite{Costa:2011mg}
\begin{equation}\label{eq:JJgeneral}
    \begin{aligned}
         \frac{\<\overline{\Ocal}_Q(y_f) \mspace{1mu} J^\mu(y_1) \mspace{1mu} J^\nu(y_2) \mspace{1mu} \Ocal_Q(y_i)\>}{\<\overline{\Ocal}_Q(y_f) \mspace{1mu} \Ocal_Q(y_i)\>} &= \frac{1}{y_{12}^{2d}} \bigg( g_0(u,v) H_{12}^{\mu\nu} + g_1(u,v) V^\mu_{1,fi} V^\nu_{2,fi} + g_2(u,v) V^\mu_{1,fi} V^\nu_{2,1i} \\
         & \hspace{1.5cm} + \, g_3(u,v) V^\mu_{1,f2} V^\nu_{2,fi} + g_4(u,v) V^\mu_{1,f2} V^\nu_{2,1i} \bigg),
    \end{aligned}
\end{equation}
where the tensor structures are defined as
\be
H_{ab}^{\mu\nu} \equiv y_{ab}^2 \mspace{1mu} g^{\mu\nu} - 2y_{ab}^\mu \mspace{1mu} y_{ab}^\nu, \qquad V^\mu_{a,bc} \equiv \frac{y_{ab}^2 \mspace{1mu} y_{ac}^2}{y_{bc}^2} \bigg( \frac{y_{ab}^\mu}{y_{ab}^2} - \frac{y_{ac}^\mu}{y_{ac}^2} \bigg),
\ee
and the various coefficients $g_a(u,v)$ are theory-dependent functions of the conformal invariant cross-ratios
\begin{equation}
    u=\frac{y_{f1}^2 \mspace{1mu} y_{2i}^2}{y_{f2}^2 \mspace{1mu} y_{1i}^2}, \qquad v=\frac{y_{fi}^2 \mspace{1mu} y_{12}^2}{y_{f2}^2 \mspace{1mu} y_{1i}^2}.
\end{equation}

We can determine the functions $g_a(u,v)$ by taking the limit $y_i \to 0$, $y_f \to \infty$ of the general four-point function in eq.~\eqref{eq:JJgeneral} and matching it to our semiclassical result in eq.~\eqref{eq:JJplane}. In this limit, the tensor structures and cross-ratios take the simpler form
\be
V^\mu_{a,fi} \to -y_{a}^\mu, \quad V^\mu_{1,f2} \to -y_{12}^\mu, \quad V^\mu_{2,1i} \to -\frac{y_{12}^2 \mspace{1mu} y_{2}^2}{y_{1}^2} \bigg( \frac{y_{12}^\mu}{y_{12}^2} + \frac{y_{2}^\mu}{y_{2}^2} \bigg), \quad u \to \frac{y_{2}^2}{y_{1}^2}, \quad v \to \frac{y_{12}^2}{y_{1}^2}.
\ee
From this limiting behavior, we can rewrite eq.~\eqref{eq:JJplane} in the manifestly conformally covariant form
\be
    \frac{\<\overline{\Ocal}_Q(\infty) \, J^\mu(y_1) J^\nu(y_2) \, \Ocal_Q(0)\>}{\<\overline{\Ocal}_Q(\infty) \Ocal_Q(0)\>} = - \bigg( \frac{Q}{\Omega_{d-1}} \bigg)^2 \frac{1}{y_{12}^{2d}} \frac{v^d}{u^{\frac{d}{2}}} V_{1,fi}^\mu V_{2,fi}^\nu,
\ee
from which we can read off the functions
\be
g_1(u,v) = \frac{v^d}{u^{\frac{d}{2}}}, \quad g_0(u,v) = g_2(u,v) = g_3(u,v) = g_4(u,v) = 0. 
\ee
We can now evaluate this Euclidean four-point function for any location of the external operators,
\be
    \frac{\<\overline{\Ocal}_Q(y_f) \mspace{1mu} J^\mu(y_1) \mspace{1mu} J^\nu(y_2) \mspace{1mu} \Ocal_Q(y_i)\>}{\<\overline{\Ocal}_Q(y_f) \mspace{1mu} \Ocal_Q(y_i)\>} = - \bigg( \frac{Q}{\Omega_{d-1}} \bigg)^2 \frac{y_{fi}^{2(d-2)}}{y_{f2}^{d-2} y_{1i}^{d-2} y_{f1}^{d-2} y_{2i}^{d-2}} \bigg( \frac{y_{f1}^\mu}{y_{f1}^2} + \frac{y_{1i}^\mu}{y_{1i}^2} \bigg) \bigg( \frac{y_{f2}^\mu}{y_{f2}^2} + \frac{y_{2i}^\mu}{y_{2i}^2} \bigg).
\ee
While this result for the leading behavior is simple, corresponding to the product of three-point functions $\<\overline{\Ocal}_Q J^\mu \Ocal_Q\> \<\overline{\Ocal}_Q J^\nu \Ocal_Q\>$, this same approach of decomposition in terms of tensor structures can be applied to more complicated expressions arising from subleading corrections in the large charge expansion, allowing us to readily translate Euclidean correlators on the cylinder to Lorentzian correlators on the plane.

We can also extend this procedure to Euclidean correlators with $k$ insertions of the conserved current, with the similar semiclassical result at large $Q$,
\be
    \frac{\<\overline{\Ocal}_Q(\infty) \mspace{1mu} J^{\mu_1}(y_1) \mspace{1mu} \cdots \mspace{1mu} J^{\mu_k}(y_k) \mspace{1mu} \Ocal_Q(0)\>}{\<\overline{\Ocal}_Q(\infty) \mspace{1mu} \Ocal_Q(0)\>} = (-1)^{\frac{k}{2}} \bigg( \frac{Q}{\Omega_{d-1}} \bigg)^k \frac{y_1^{\mu_1}}{y_1^d} \cdots \frac{y_k^{\m_k}}{y_k^d}.
\ee
While the set of possible tensor structures is more complicated for higher $k$, for the leading behavior as $Q \to \infty$ we again find that only one combination contributes to this correlator,
\be
    \frac{\<\overline{\Ocal}_Q(\infty) \mspace{1mu} J^{\mu_1}(y_1) \mspace{1mu} \cdots \mspace{1mu} J^{\mu_k}(y_k) \mspace{1mu} \Ocal_Q(0)\>}{\<\overline{\Ocal}_Q(\infty) \mspace{1mu} \Ocal_Q(0)\>} \propto V_{1,fi}^{\mu_1} \mspace{1mu} \cdots \mspace{1mu} V_{k,fi}^{\mu_k},
\ee
allowing us to reconstruct this correlation function for arbitrary $y_i$ and $y_f$.

Finally, we can consider correlation functions involving the stress-energy tensor, such as the four-point function
\be
    \frac{\<\overline{\Ocal}_Q(\infty) \mspace{1mu} T^{\mu\nu}(y_1) \mspace{1mu} T^{\rho\sigma}(y_2) \mspace{1mu} \Ocal_Q(0)\>}{\<\overline{\Ocal}_Q(\infty) \mspace{1mu} \Ocal_Q(0)\>} = \bigg( \frac{d\D_Q}{(d-1)\Omega_{d-1}} \bigg)^2 \bigg(\frac{y_1^\mu \mspace{1mu} y_1^\nu}{y_1^{2d+2}} - \frac{1}{d} \frac{g^{\mu\nu}}{y_1^{2d}} \bigg) \bigg(\frac{y_2^\rho \mspace{1mu} y_2^\sigma}{y_2^{2d+2}} - \frac{1}{d} \frac{g^{\rho\sigma}}{y_2^{2d}} \bigg).
\ee
We can again write down all possible tensor structures built from $H_{ab}^{\mu\nu}$ and $V_{a,bc}^\mu$ and match that general expression to this particular correlator. Fortunately, just like for the $U(1)$ current, this semiclassical expression factorizes into a product of three-point functions, allowing us to obtain the simple result
\be
    \begin{aligned}
        &\frac{\<\overline{\Ocal}_Q(y_f) \mspace{1mu} T^{\mu\nu}(y_1) \mspace{1mu} T^{\rho\sigma}(y_2) \mspace{1mu} \Ocal_Q(y_i)\>}{\<\overline{\Ocal}_Q(y_f) \mspace{1mu} \Ocal_Q(y_i)\>} \\
        & \qquad = \bigg( \frac{d\D_Q}{(d-1)\Omega_{d-1}} \bigg)^2 \frac{1}{y_{12}^{2d+2}} \frac{v^{d+1}}{u^{\frac{d+1}{2}}} \bigg(V_{1,fi}^\mu V_{1,fi}^\nu - \frac{1}{d} V_{1,fi}^2 \mspace{1mu} g^{\mu\nu} \bigg) \bigg(V_{2,fi}^\rho V_{2,fi}^\sigma - \frac{1}{d} V_{2,fi}^2 \mspace{1mu} g^{\rho\sigma} \bigg),
    \end{aligned}
\ee
with an analogous expression for $k$ insertions of the stress tensor.


\bibliographystyle{utphys}
\bibliography{draft}{}

\end{document}